\documentclass[preprintnumbers]{revtex4}
\usepackage{amsfonts}
\usepackage{amssymb}
\usepackage{graphicx}
\usepackage{subfigure}
\usepackage{color}
\usepackage{amsmath}

\input{tcilatex}
\begin{document}

\title{Integer and half-integer channel spins for elastic scattering cross
sections}
\author{A. S. Tkachenko,$^{1,2}$ R. Ya. Kezerashvili,$^{3,4}$ N. A. Burkova,$%
^{2}$ S. B. Dubovichenko$^{1,2}$}
\affiliation{$^{1}$Fesenkov Astrophysical Institute \textquotedblleft
NCSRT\textquotedblright\ ASA MDASI RK, 050020, Almaty, Kazakhstan\\
$^{2}$al-Farabi Kazakh National University, 050040, Almaty, Kazakhstan\\
$^{3}$New York City College of Technology, City University of New York,
Brooklyn, NY 11201, USA\\
$^{4}$Graduate School and University Center, City University of New York,
New York 10016, USA}
\date{\today }

\begin{abstract}
We present the analytical expressions for the differential cross sections
and independent partial amplitudes for elastic scattering of nuclear
particles for channels with a spin value of 1/2, 1, 3/2, 2 and 5/2. The
independent partial amplitudes are presented for arbitrary orbital angular
momentum $l$ and taking into account spin-orbit splitting. The analytical
expressions allow one to carry out full phase shift analyses using
experimental data for differential cross sections for processes with channel
spins 1/2, 1, 3/2, 2 and 5/2.
\end{abstract}

\maketitle

\section{Introduction}

One of the most important problems in the theory of nuclear reactions and
nuclear scattering is a construction of potentials for a nucleon-nucleon,
nucleon-nuclear and nuclear-nuclear interactions. These potentials are used
for calculation of any nuclear characteristics, for example, the binding
energies of the nuclei, the properties of their bound states, cross
sections, astrophysical $S-$factors or rates of various reactions including
thermonuclear processes at ultra low energies \cite{Dubovichenko2019}. Today
the known nucleon-nucleon interactions are widely used for studies of $N-d,$
$N-^{3}$He, $N-^{3}$H and $N-^{4}$He processes within different few-body
methods. The exact treatment of nuclei starting from the constituent
nucleons and the fundamental interactions among them has been a
long-standing goal in nuclear physics \cite{Quaglioni}. In the case of
collision of particles composed of identical protons and identical neutrons
the effective interactions should be constructed which, in principle, change
whenever colliding particles are different or methods of construction are
used. However, in general the effective interaction are energy-dependent and
complex. The scattering theory allows the calculation of the scattering
cross sections as a function of the scattering angle and energy for any
potential. In the case of nuclear reactions, the potential of interaction
between two colliding particles \textit{a priori} is unknown. Thus, the task
is to get as much information as possible related to the interaction
potential through a systematic analysis of the experimental data on the
scattering cross section, as well as polarization measurements. The problem
of finding the nucleon-nuclear or nuclear-nuclear potential on the basis of
the obtained experimental cross sections is twofold: 1) the determination of
the phase shifts from obtained experimental scattering cross sections; 2)
the construction of the nucleon-nuclear or nuclear-nuclear potential based
on the obtained phase shifts of elastic scattering processes \cite%
{Friedrich2013,Henley2007,Hodgson1963,NucPhys2019}.

Many nuclear processes of formation of light elements are results of the
radiative capture of protons or neutrons by light nuclei as well as
collisions of light nuclei. These processes are parts of the nucleosynthesis
chain of light elements in the primordial Universe and occur in the early
stage of stable star formation \cite{Barnes,Iocco2009}. A study of these
processes requires a construction of nucleon-nuclear and nuclear-nuclear
potentials based on the description of the known scattering phase shifts and
the main characteristics of the ground and excited states of nuclei.
However, the direct measurement of cross sections of the charged-particle
induced reactions at thermonuclear energies becomes almost impossible due to
its slump with energy decreasing because of the Coulomb barrier. A common
way to get the astrophysical $S-$factors in the low-energy region (stellar
energies) is its extrapolation based on the trend of experimental data in
the high-energy region \cite{Fang2016}. However, prior research \cite{He2013}
demonstrates the danger of simply extrapolating experimental data over a
very large energy range towards stellar energies. It also shows the
necessity for carefully studying the low-energy reaction cross sections. The
predicted light element abundance highly depends on the input data such as
the nuclear reaction rates, the equation of state, the opacity of the
stellar matter. Nuclear astrophysics reaction rate determines the
possibility of a particular reaction, thereby affecting the process of
stellar evolution and elements formation. In reaction network calculations,
all reactions related to formation and destruction of light nuclei should be
taken into account in order to get accurate results \cite{Coc2012}. In
particular, this related to $^{6}$Li and $^{7}$Li abundance \cite%
{Anders2014,Hiyama2018,Gnech2019,NucPhys2019}. Therefore, a careful
theoretical analysis of astrophysical reactions at low energies requires the
corresponding nucleon-nuclear and nuclear-nuclear potentials.

Phase shift analysis is usually performed on the basis of expressions for
the elastic scattering cross sections of nuclear particles, namely, on the
basis of the representation of the differential cross sections for the
elastic scattering through the phase shifts of elastic scattering \cite%
{Nichitiu1981,Nichitiu1980,Dubovichenko2015}. Many of these expressions can
be found in the scientific literature, and were subsequently collected in
books \cite{Satchler2009} and \cite{Dubovichenko2015}. For example, the
scattering of spinless particles on a target with also zero spin and the
system of non-identical particles with a channel\ spin $S=1/2$, such as $%
N-^{4}$He, is described in detail in the books \cite%
{Hodgson1963,Dubovichenko2015}. However, a particular difficulty is the
calculation of differential cross sections for high-spin states (channel
spin $S>1$) of a system of particles. Results for a high channel spin, in
particular, for a particle system with the spin structure $1/2+1$ and $%
1/2+3/2$ a high channel spin were given earlier in Refs. \cite{Seyler1969}
and \cite{Brown1973}, respectively. However, we still lack fully analytical
expressions for calculations of the differential elastic scattering cross
sections for half-integer channel spins\ $S=3/2$ and $S=5/2$ and integer
channel spins $S=1$ and $S=2$. We already reported preliminary calculations
related to these channel spins \cite{Tkachenko}. This article presents the
final analytical expressions for calculations of the differential cross
sections for the elastic scattering of non-identical nuclear particles. All
differential cross sections are expressed in the terms of the orbital
quantum number $l$, which allows one to explicitly take into account in the
calculations a certain energy-dependent number of scattering partial waves
and to consider the relative contribution of each partial wave. These
expressions allow one to perform the phase shift analyses based on
experimental cross sections obtained for nuclear-nuclear scattering for the
integer and half-integer channel spin of the spin-$1/2-$spin-$3/2$, spin-$%
1/2-$ spin-$1$, and spin-$1/2-$spin-$2$ and spin-$1-$spin-$3/2$ processes.
The amplitudes obtained for particular channel spins, along with the data on
elastic channels, make it possible to carry out full phase shift analysis
with channel coupling. Analyses and calculations of cross sections and
reactions rates at astrophysical energies require a small number of partial
waves. In this case from general formulas presented in this work one obtains
simple and convenient algebraic expressions for the partial amplitudes.

This work is organized as follows: in Sec. \ref{Crosssection} we present the
main ingredients for the description of differential cross sections.
Processes with spin-$1/2-$spin-$3/2$, spin-$1/2-$ spin-$1$, and spin-$1/2-$%
spin-$2$ and spin-$1-$spin-$3/2$ are considered in Secs. \ref{Halfinteger}, %
\ref{Spin1} and \ref{Spin2}, respectively. The analytical expressions for
the independent partial amplitudes for above mentioned processes are given
in Appendixes A, B and C, correspondingly. Conclusions follow in Sec. \ref%
{Conclusions}.

\section{\protect\bigskip Theoretical formalism for particles with spins}

\label{Crosssection}

We consider in general a collision of two particles with spins. One of the
particles can be a nucleon or the both particles are nuclei. When the two
colliding particles have spins, these spins may be reoriented by the
scattering even if internal excitations of nuclei do not occur. The total
angular momentum and its projection are conserved so that flipping one spin,
which is related to the changing the value of its projection on the $z-$axis
is compensated by a flip of the other or by an exchange of angular momentum
with the relative orbital motion. In the latter case, the plane of the
orbits is tilted.

The differential cross section of the elastic scattering channel can be
written as \cite{Satchler2009}
\begin{equation}
\frac{d\sigma }{d\Omega }=\underset{S\nu \nu ^{\prime }}{\sum }g(S)\frac{%
d\sigma \left( S\right) }{d\Omega },  \label{Generalsigma}
\end{equation}%
where $g(S)$ is the statistical weight for channel spin $S$

\begin{equation}
g(S)=\frac{2S+1}{\left( 2s_{1}+1\right) \left( 2s_{2}+1\right) }.
\label{g(S)}
\end{equation}%
The spin of the scattering reaction channel is the result of the vector
addition of the spins $\mathbf{s}_{1}$ and $\mathbf{s}_{2}$ of the incident
and target particles and is

\begin{equation}
\mathbf{S}=\mathbf{s}_{1}+\mathbf{s}_{2},~\left\vert s_{1}-s_{2}\right\vert
\leq S\leq s_{1}+s_{2}.  \label{channelspin}
\end{equation}%
In general case the partial differential cross sections can be written as

\begin{equation}
\frac{d\sigma (S)}{d\Omega }=\underset{SS^{^{\prime }}\nu \nu ^{\prime }}{%
\sum }g(S)M_{S^{\prime }\nu ^{\prime }}^{S\nu ^{\ast }}M_{S^{\prime }\nu
^{\prime }}^{S\nu },  \label{partial-cross-section}
\end{equation}%
where $M_{S^{\prime }\nu ^{\prime }}^{S\nu }$ are the matrix elements of the
transition $M-$matrix. Here we assume the general case when the potential
couples spins of both collided particles. For the elastic scattering when $%
S=S^{\prime }$, Eq. (\ref{partial-cross-section}) describes the partial
cross section in Eq. (\ref{Generalsigma}). In the channel spin
representation, the relationship between the transition $M-$matrix and the
scattering or collision matrix $U$ in the general case, when the initial and
final channel spins $S$ and $S^{\prime }$ can take any possible values, is
described by the expansion of the scattering amplitude in terms of the
Legendre polynomials $P_{n}^{m}(\theta )$ as \cite{Lane1958}:

\begin{eqnarray}
M_{S^{\prime }\nu ^{\prime }}^{S\nu }(\theta ) &=&f(\theta )\delta
_{SS^{\prime }}\delta _{\nu \nu ^{\prime }}+\frac{i}{2k}\underset{%
Jll^{\prime }}{\dsum }\sqrt{\frac{\left( l^{\prime }-m^{\prime }\right) !}{%
\left( l^{\prime }+m^{\prime }\right) !}}C_{S\nu l0}^{J\nu }C_{S^{\prime
}\nu ^{\prime }l^{\prime }m^{\prime }}^{J\nu }  \label{MUrelation3} \\
&&\left. \times \exp (i(\omega _{l}+\omega _{l^{\prime }}))\left( \delta
_{SS^{\prime }}\delta _{ll^{\prime }}-U_{S^{\prime }l^{\prime }Sl}^{J\pi
}\right) P_{l^{\prime }}^{m^{\prime }}(\theta )\right] .  \notag
\end{eqnarray}%
In Eq. (\ref{MUrelation3}) the following notations are used: $S$ and $%
S^{\prime }$ are the spins of the initial and final channels, respectively, $%
\nu $ and $\nu ^{\prime }$ are their projections on $z-$ axis, $l$ and $%
l^{\prime }$ are the orbital quantum numbers of the initial and final
channels, $C_{a\alpha b\beta }^{c\gamma }$ are the Clebsch-Gordan
coefficients, $\omega _{l}$ are the Coulomb scattering phase shifts, and $%
f(\theta )$ and $U_{S^{\prime }l^{\prime }Sl}^{J\pi }$ are the Coulomb
scattering amplitude and the scattering matrix, respectively. The Coulomb
amplitude is defined as follows

\begin{equation}
f_{c}\left( \theta \right) =-\left( \frac{\gamma }{2k\sin ^{2}\left( \theta
/2\right) }\right) \exp \left( i\gamma \ln \left[ \sin ^{-2}\left( \theta
/2\right) \right] +2i\omega _{0}\right) .  \label{Coulombscattering}
\end{equation}%
In Eq. (\ref{Coulombscattering}) $\theta $ is the scattering angle, $k$ is
the wave number of the relative motion of collided particles, $k^{2}=\frac{%
2\mu E}{\hbar ^{2}},~$where $\mu $ is the reduced mass of particles and $E$
is the energy of colliding particles in the center-of-mass system, $\gamma =%
\frac{\mu Z_{1}Z_{2}e^{2}}{k\hbar ^{2}}$ is the Coulomb parameter, where $Z$
is the particle's charge in units of the elementary charge. For a proper
description of the scattering cross section, it is sufficient to take into
account only elastic channels. Then the collision $U-$matrix takes the form:

\begin{equation}
U_{Sl}=\exp \left( 2i\delta _{Sl}\right) .  \label{Umatrix}
\end{equation}

The description of partial differential cross sections requires only those
amplitudes for which $S=S^{\prime }$. The matrix elements with $S\neq
S^{\prime }$ can be used for calculations of polarization effects, which
require the consideration of spin-mixing states. For analytical calculations
of elastic scattering cross sections i.e. for the case when the spins of the
initial and final channels have the same value $\left( S=S^{\prime }\right) $
it is convenient to present the partial amplitudes of the $M-$matrix in a
more succinct form. For this case the expression (\ref{MUrelation3}) could
be written in the following form

\begin{equation}
\overset{\sim }{M}_{\text{ }\nu ^{\prime }}^{S\nu }(\theta )\equiv M_{S\nu
^{\prime }}^{S\nu }(\theta )=\underset{Jlm}{\dsum }C_{S\nu l0}^{J\nu
}C_{S\nu ^{\prime }lm}^{J\nu }A_{Jl}^{S}Y_{lm}\left( \theta ,0\right) ,
\label{M_compact}
\end{equation}%
where

\begin{equation}
A_{Jl}^{S}(\theta )=\frac{\sqrt{\pi }}{k}\left\{
\begin{array}{c}
f_{c}\left( \theta \right) +i\left( 2l+1\right) ^{1/2}\exp \left( 2i\omega
_{l}\right) \left( 1-U_{Sl}^{J}\right) ,\text{ \ }if\text{ \ \ \ }\nu =\nu
^{\prime }; \\
i\left( 2l+1\right) ^{1/2}\exp \left( 2i\omega _{l}\right) \left(
1-U_{Sl}^{J}\right) ,\text{ \ \ \ \ \ \ \ \ \ \ \ \ \ }if\text{ \ \ \ }\nu
\neq \nu ^{\prime }.%
\end{array}%
\right.  \label{Elem1}
\end{equation}%
The next step consists in recoupling procedure for the Clebsch-Gordan
coefficients in (\ref{M_compact}) using the relevant expression from \cite%
{Varshalovich2017}

\begin{equation}
C_{a\alpha f\varphi }^{d\delta }C_{b-\beta e\varepsilon }^{d\delta }=%
\underset{c\gamma }{\sum }\left( -1\right) ^{c+d-\beta -\varphi }\left(
2d+1\right) C_{a\alpha b\beta }^{c\gamma }C_{f-\varphi e\varepsilon
}^{c\gamma }\left\{
\begin{array}{c}
a~b~c \\
e~f~d%
\end{array}%
\right\} ,  \label{Varshalovich1}
\end{equation}%
where $\left\{
\begin{array}{c}
... \\
...%
\end{array}%
\right\} $ are Wigner $6j-$symbols. Therefore, using one-to-one
correspondence for the $\left\{ afd\right\} \rightarrow \left\{ SlJ\right\} $
and $\left\{ bed\right\} \rightarrow \left\{ SlJ\right\} $ we can transform
the product of the Clebsch -- Gordan coefficients as

\begin{equation}
C_{S\nu l0}^{J\nu }C_{S\nu ^{\prime }lm}^{J\nu }=\underset{\kappa m}{\dsum }%
\left( -1\right) ^{\kappa +J-\nu ^{\prime }}\left( 2J+1\right) C_{S\nu
S-v^{\prime }}^{\kappa m}C_{l0lm}^{\kappa m}\left\{
\begin{array}{c}
SS\kappa \\
llJ%
\end{array}%
\right\} .  \label{Elem2}
\end{equation}%
According to the rules of addition of angular momenta, the summation by $%
\kappa $ in Eq. (\ref{Elem2}) includes the integers in the intervals $0\leq
\kappa \leq 2S$ and $0\leq \kappa \leq 2l.$

Using the following symmetry properties of the Clebsch-Gordan coefficients
\cite{Varshalovich2017}

\begin{eqnarray}
C_{a\alpha b\beta }^{c\gamma } &=&\left( -1\right) ^{a+b-c}C_{b\beta a\alpha
}^{c\gamma },  \notag \\
C_{a\alpha b\beta }^{c\gamma } &=&\left( -1\right) ^{b+\beta }\sqrt{\frac{%
2c+1}{2a+1}}C_{c-\gamma b\beta }^{a-\alpha },  \notag \\
C_{a\alpha b\beta }^{c\gamma } &=&\left( -1\right) ^{a+b-c}C_{a-\alpha
b-\beta }^{c-\gamma },  \label{Varshalovich2}
\end{eqnarray}%
the first coefficient $C_{S\nu S-v^{\prime }}^{\kappa m}$ in Eq. (\ref{Elem2}%
) can be rewritten as

\begin{equation}
C_{S\nu S-\nu ^{\prime }}^{\kappa m}=\left( -1\right) ^{S+\kappa +\nu
^{\prime }}\sqrt{\frac{2\kappa +1}{2S+1}}C_{\kappa mS\nu ^{\prime }}^{S\nu }.
\label{Elem3}
\end{equation}%
Thus, Eqs. (\ref{Elem1}), (\ref{Elem2}) and (\ref{Elem3}) allow one to
obtain the following form of the matrix elements $\overset{\sim }{M}_{\text{
}\nu ^{\prime }}^{S\nu }(\theta )$:

\begin{equation}
\overset{\sim }{M}_{\text{ }\nu ^{\prime }}^{S\nu }(\theta )=\underset{%
\kappa Jlm}{\dsum }\left( -1\right) ^{J+S}\left( 2J+1\right) \sqrt{\frac{%
2\kappa +1}{2S+1}}C_{\kappa mS\nu ^{\prime }}^{S\nu }C_{l0lm}^{\kappa
m}\left\{
\begin{array}{c}
SS\kappa \\
llJ%
\end{array}%
\right\} A_{Jl}^{S}Y_{lm}\left( \theta ,0\right) .  \label{Elem4}
\end{equation}%
Let us turn now to the construction of the differential cross section basing
on the $\overset{\sim }{M}_{\text{ }\nu ^{\prime }}^{S\nu }$ matrix
elements. According to (\ref{Generalsigma}) the differential cross section
can be written as

\begin{equation}
\frac{d\sigma (S)}{d\Omega }=\underset{S\nu \nu ^{\prime }}{\sum }g(S)%
\overset{\sim }{M}_{\text{ }\nu ^{\prime }}^{S\nu ^{\ast }}\overset{\sim }{M}%
_{\text{ }\nu ^{\prime }}^{S\nu }.  \label{elastic_cross_section}
\end{equation}%
Substituting Eq. (\ref{Elem4}) into (\ref{elastic_cross_section}) we can
perform the external summation over $\nu \nu ^{\prime }$ using the
well-known relation for the Clebsch-Gordan coefficients \cite%
{Varshalovich2017}

\begin{equation}
\underset{a\alpha }{\sum }C_{a\alpha b\beta }^{c\gamma }C_{a\alpha b\beta
^{\prime }}^{c\gamma ^{\prime }}=\frac{\Pi _{c}^{2}}{\Pi _{b}^{2}}\delta
_{bb^{\prime }}\delta _{\beta \beta ^{\prime },}\text{\ \ where \ }\Pi
_{ab...c}=\sqrt{\left( 2a+1\right) \left( 2b+1\right) ...\left( 2c+1\right) }%
.\   \label{Varshalovich4}
\end{equation}%
The application of (\ref{Varshalovich4}) leads to

\begin{equation}
\sum\limits_{vv^{^{\prime }}}C_{\kappa mS\nu ^{\prime }}^{S\nu }C_{\kappa
^{^{\prime }}m^{^{\prime }}S\nu ^{\prime }}^{S\nu }=\delta _{\kappa \kappa
^{\prime }}\delta _{mm^{\prime }}\frac{2S+1}{2\kappa +1}.\   \label{Elem5}
\end{equation}
Thus, considering (\ref{Elem4}) and (\ref{Elem5}) in Eq. (\ref%
{elastic_cross_section}), we obtain the final expression for the
differential cross section as follows:

\begin{equation}
\frac{d\sigma }{d\Omega }=\underset{S\kappa }{\dsum }g\left( S\right)
\left\vert \underset{Jlm}{\dsum }\left( -1\right) ^{J}\left( 2J+1\right)
C_{l0lm}^{\kappa m}\left\{
\begin{array}{c}
SS\kappa \\
llJ%
\end{array}%
\right\} A_{Jl}^{S}Y_{lm}\left( \theta ,0\right) \right\vert ^{2}.
\end{equation}%
It should be noted that this expression for the differential cross section
has a universal form and can be applied to any processes of elastic
scattering, regardless of whether the channel spins are integer or
half-integer.

In Table \ref{tab1} are listed the light nuclei with the spins $1/2$, $1$, $%
3/2$ and $2$ relevant to the nucleosynthesis chain of light elements in the
primordial Universe.

\begin{table}[b]
\caption{Possible incident particles and targets. $J^{\protect\pi }$ is the
angular momentum and parity.}
\label{tab1}
\begin{center}
\begin{tabular}{cccccccccccccccccc}
\hline
$Z$ & $0$ & $1$ & $1$ & $1$ & $1$ & $2$ & $3$ & $3$ & $3$ & $3$ & $3$ & $3$
& $3$ & $4$ & $4$ & $4$ & $5$ \\ \hline
Particle & $n$ & $p$ & d & $^{3}$H & $^{4}$H & $^{3}$He & $^{3}$Li & $^{4}$Li
& $^{5}$Li & $^{6}$Li & $^{7}$Li & $^{9}$Li & $^{10}$Li & $^{7}$Be & $^{9}$Be
& $^{11}$Be & $^{7}$B \\ \hline
$J^{\pi }$ & $1/2^{+}$ & $1/2^{+}$ & $1^{+}$ & $1/2^{+}$ & $2^{-}$ & $%
1/2^{+} $ & $2^{+}$ & $2^{-}$ & $3/2^{-}$ & $1^{+}$ & $3/2^{-}$ & $3/2^{-}$
& $1^{-}$ & $3/2^{-}$ & $3/2^{-}$ & $1/2^{+}$ & $3/2^{-}$ \\ \hline\hline
$Z$ & $5$ & $5$ & $5$ & $5$ & $5$ & $5$ & $5$ & $6$ & $6$ & $6$ & $6$ & $6$
& $6$ & $7$ & $7$ & $7$ & $7$ \\ \hline
Particle & $^{8}$B & $^{9}$B & $^{11}$B & $^{12}$B & $^{14}$B & $^{15}$B & $%
^{17}$B & $^{9}$C & $^{11}$C & $^{13}$C & $^{15}$C & $^{17}$C & $^{19}$C & $%
^{10}$N & $^{13}$N & $^{15}$N & $^{16}$N \\ \hline
$J^{\pi }$ & $2^{+}$ & $3/2^{-}$ & $3/2^{-}$ & $1^{+}$ & $3/2^{-}$ & $2^{-}$
& $3/2^{-}$ & $3/2^{-}$ & $3/2^{-}$ & $1/2^{-}$ & $1/2^{+}$ & $3/2^{+}$ & $%
1/2^{+}$ & $2^{-}$ & $1/2^{-}$ & $1/2^{-}$ & $2^{-}$ \\ \hline
\end{tabular}%
\end{center}
\end{table}
\bigskip

\section{Processes with spin-$1/2$ $-$ spin-$3/2$}

\label{Halfinteger}

Let us consider the spin states for the collision of two particles with spin
$1/2$ and $3/2.$ In this case the collision of following particles from
Table \ref{tab1} can be considered: $n,$ $p,$ $^{3}$H(1/2$^{+}$)$,^{3}$He(1/2%
$^{+}$) as the first particle and $^{5}$Li(3/2$^{-}$), $^{7}$Li(3/2$^{-}$), $%
^{9}$Li(3/2$^{-}$), $^{7}$Be(3/2$^{-}$), $^{9}$Be(3/2$^{-}$), $^{7}$B(3/2$%
^{-}$), $^{9}$B(3/2$^{-}$), $^{11}$B(3/2$^{-}$), $^{13}$B(3/2$^{-}$), $^{15}$%
B(3/2$^{-}$), $^{9}$C(3/2$^{-}$), $^{11}$C(3/2$^{-}$), $^{17}$C(3/2$^{+}$), $%
^{13}$O(3/2$^{-}$) as the second one.

In this case channel spin can be $S=1$ and $S=2$ according to the vector
addition (\ref{channelspin}). For the channel with the integer spin values 1
and 2, the $M-$matrix is represented as

\begin{eqnarray}
&&%
\begin{array}{cccccccccc}
S\text{ }\upsilon &  & \text{ \ \ \ }22 & \text{ \ \ \ \ \ \ }21 & \text{ \
\ \ \ \ }20 & \text{ \ \ }2-1 & \text{ \ }2-2 & \text{ \ \ \ \ }11 & \text{
\ \ \ \ \ }10 & \text{ \ \ \ }1-1 \\
S^{^{\prime }}\text{ }\upsilon ^{^{\prime }} &  &  &  &  &  &  &  &  &
\end{array}
\notag \\
M &=&%
\begin{array}{cc}
22 &  \\
21 &  \\
20 &  \\
2-1 &  \\
2-2 &  \\
11 &  \\
10 &  \\
1-1 &
\end{array}%
\left[
\begin{array}{cccccccc}
\mathfrak{Q}_{\text{\ \ }2}^{22} & \mathfrak{Q}_{\text{\ \ }2}^{21} &
\mathfrak{Q}_{\text{\ \ }2}^{20} & -\mathfrak{Q}_{-2}^{21} & \mathfrak{Q}_{%
\text{ \ \ \ }2}^{2-2} & \mathfrak{M}_{22}^{11} & \mathfrak{M}_{22}^{10} &
\mathfrak{M}_{2\text{ \ }2}^{1-1} \\
\mathfrak{Q}_{\text{ \ }1}^{22} & \mathfrak{Q}_{\text{ \ }1}^{21} &
\mathfrak{Q}_{\text{ \ }1}^{20} & \mathfrak{Q}_{\text{ \ \ \ }1}^{2-1} & -%
\mathfrak{Q}_{-1}^{22} & \mathfrak{M}_{21}^{11} & \mathfrak{M}_{21}^{10} &
\mathfrak{M}_{2\text{ \ }1}^{1-1} \\
\mathfrak{Q}_{\text{ \ }0}^{22} & \mathfrak{Q}_{\text{ \ }0}^{21} &
\mathfrak{Q}_{\text{ \ }0}^{20} & -\mathfrak{Q}_{\text{ \ }0}^{21} &
\mathfrak{Q}_{\text{ \ }0}^{22} & \mathfrak{M}_{20}^{11} & 0 & \mathfrak{M}%
_{20}^{11} \\
\mathfrak{Q}_{-1}^{22} & \mathfrak{Q}_{\text{ \ \ \ }1}^{2-1} & -\mathfrak{Q}%
_{\text{ \ }1}^{20} & \mathfrak{Q}_{\text{ \ }1}^{21} & -\mathfrak{Q}_{\text{
\ }1}^{22} & -\mathfrak{M}_{2\text{ \ }1}^{1-1} & \mathfrak{M}_{21}^{10} & -%
\mathfrak{M}_{21}^{11} \\
\mathfrak{Q}_{\text{ \ \ \ }2}^{2-2} & \mathfrak{Q}_{-2}^{21} & \mathfrak{Q}%
_{\text{\ \ }2}^{20} & -\mathfrak{Q}_{\text{\ \ }2}^{21} & \mathfrak{Q}_{%
\text{\ \ }2}^{22} & \mathfrak{M}_{2\text{ \ }2}^{1-1} & -\mathfrak{M}%
_{22}^{10} & \mathfrak{M}_{22}^{11} \\
\mathfrak{M}_{11}^{22} & \mathfrak{M}_{11}^{21} & \mathfrak{M}_{11}^{20} &
\mathfrak{M}_{1\text{ \ }1}^{2-1} & \mathfrak{M}_{1\text{ \ }1}^{2-2} &
\mathcal{T}_{\text{ \ }1}^{11} & \mathcal{T}_{\text{ \ }1}^{10} & \mathcal{T}%
_{\text{ \ \ \ }1}^{1-1} \\
\mathfrak{M}_{10}^{22} & \mathfrak{M}_{10}^{21} & 0 & \mathfrak{M}_{10}^{21}
& -\mathfrak{M}_{10}^{22} & \mathcal{T}_{\text{\ \ }0}^{11} & \mathcal{T}_{%
\text{ \ }0}^{10} & -\mathcal{T}_{\text{\ \ }0}^{11} \\
\mathfrak{M}_{1\text{ \ }1}^{2-2} & -\mathfrak{M}_{1\text{ \ }1}^{2-1} &
\mathfrak{M}_{11}^{20} & -\mathfrak{M}_{11}^{21} & \mathfrak{M}_{11}^{22} &
\mathcal{T}_{\text{ \ \ \ }1}^{1-1} & -\mathcal{T}_{\text{ \ }1}^{10} &
\mathcal{T}_{\text{ \ }1}^{11}%
\end{array}%
\right] .  \label{integermatrix}
\end{eqnarray}%
In matrix (\ref{integermatrix}) we use the calligraphic letter $\mathcal{T}$
\ for the notation of triplet spin states for partial amplitudes and the
Fraktur letters $\mathfrak{Q}$ and $\mathfrak{M}$ have been denoted for the
quintet and spin-mixing states partial amplitudes, respectively. We
especially labeled spin-mixing states by the letter $\mathfrak{M}$ to avoid
later confusion with the general labeling of matrix elements of the $M-$%
matrix.

The matrix (\ref{integermatrix}) has in total 64 partial amplitudes.
According to the parity conservation which restricts the sum $l+l^{\prime }$
to being even, it is easy to show that

\begin{equation}
M_{S^{\prime }-v^{\prime }}^{S-v}=\left\{
\begin{array}{c}
\left( -1\right) ^{S^{\prime }-S+v^{\prime }-v}M_{S^{\prime }v^{\prime
}}^{Sv},\text{ \ }if\text{ \ \ \ }S\neq S^{\prime }, \\
\left( -1\right) ^{v^{\prime }-v}\overset{\sim }{M}_{\text{ \ }v^{\prime
}}^{Sv},\text{ \ \ \ \ \ \ \ \ \ }if\text{ \ \ \ }S=S^{\prime }.%
\end{array}%
\right.  \label{Parity}
\end{equation}%
Thus, it is possible to reduce the number of independent partial amplitudes
from 64 to 34. Considering that 2 of 16 spin-mixing partial amplitudes equal
0, finally 32 independent $M-$matrix elements remain.

The differential cross section for the elastic scattering in a system of two
particles with spin-$1/2-$ spin-$3/2$, taking into account the spin-orbit
interaction, is presented in the form:

\begin{equation}
\frac{d\sigma \left( \theta \right) }{d\Omega }=\frac{3}{8}\frac{d\sigma _{%
\mathcal{T}}}{d\Omega }+\frac{5}{8}\frac{d\sigma _{\mathfrak{Q}}}{d\Omega }.
\label{cross_sec1}
\end{equation}%
The differential cross section $d\sigma _{\mathcal{T}}/d\Omega $ corresponds
to the $triplet$ state with the channel spin $S=S^{\prime }=1$ and can be
constructed as a combination of independent partial amplitudes from the
matrix (\ref{integermatrix}). The differential cross section for the elastic
scattering in the triplet channel is determined only by five independent
partial amplitudes $\mathcal{T}_{\text{ \ }0}^{10},$ $\mathcal{T}_{\text{ \ }%
1}^{11},\mathcal{T}_{\text{ \ }1}^{10},\mathcal{T}_{\text{\ \ }0}^{11}$, and
$\mathcal{T}_{\text{ \ \ \ }1}^{1-1}$ given by $3\times 3$ matrix in the
lower corner of the matrix (\ref{integermatrix}) and can by written as

\begin{equation}
\frac{d\sigma _{\mathcal{T}}}{d\Omega }=\frac{1}{3}\left[ \left\vert
\mathcal{T}_{\text{ \ }0}^{10}\right\vert ^{2}+2\left( \left\vert \mathcal{T}%
_{\text{ \ }1}^{11}\right\vert ^{2}+\left\vert \mathcal{T}_{\text{ \ }%
1}^{10}\right\vert ^{2}+\left\vert \mathcal{T}_{\text{\ \ }%
0}^{11}\right\vert ^{2}+\left\vert \mathcal{T}_{\text{ \ \ \ }%
1}^{1-1}\right\vert ^{2}\right) \right] .  \label{triplet_cross_section}
\end{equation}%
The analytical expressions obtained for the independent partial amplitudes $%
\mathcal{T}_{\text{ \ }v^{^{\prime }}}^{1v}$ for the spin triplet state are
presented in Appendix A by Eqs. (\ref{A1A3}) - (\ref{A5E3}).

To describe a $quintet$ spin states with\ the channel spin $S=S^{\prime }=2$
13 independent amplitudes are required. The differential cross section $%
d\sigma _{\mathfrak{Q}}/d\Omega $ for the elastic scattering for the quintet
spin state is described by the expression

\begin{eqnarray}
\frac{d\sigma _{\mathfrak{Q}}}{d\Omega } &=&\frac{1}{3}\left[ \left\vert
\mathfrak{Q}_{\text{ \ }0}^{20}\right\vert ^{2}+2\left( \left\vert \mathfrak{%
Q}_{\text{ \ }1}^{21}\right\vert ^{2}+\left\vert \mathfrak{Q}_{\text{\ \ }%
2}^{22}\right\vert ^{2}+\left\vert \mathfrak{Q}_{\text{ \ }%
1}^{20}\right\vert ^{2}+\left\vert \mathfrak{Q}_{\text{ \ }%
0}^{21}\right\vert ^{2}+\left\vert \mathfrak{Q}_{\text{ \ }%
1}^{22}\right\vert ^{2}+\left\vert \mathfrak{Q}_{\text{\ \ }%
2}^{21}\right\vert ^{2}+\left\vert \mathfrak{Q}_{\text{ \ \ \ }%
1}^{2-1}\right\vert ^{2}+\right. \right.  \label{quintet} \\
&&\left. \left. +\left\vert \mathfrak{Q}_{\text{\ \ }2}^{20}\right\vert
^{2}+\left\vert \mathfrak{Q}_{\text{ \ }0}^{22}\right\vert ^{2}+\left\vert
\mathfrak{Q}_{-2}^{21}\right\vert ^{2}+\left\vert \mathfrak{Q}%
_{-1}^{22}\right\vert ^{2}+\left\vert \mathfrak{Q}_{\text{ \ \ \ }%
2}^{2-2}\right\vert ^{2}\right) \right] .  \notag
\end{eqnarray}%
The independent partial amplitudes $\mathfrak{Q}_{\text{ \ }0}^{20},$ $%
\mathfrak{Q}_{\text{ \ }1}^{21},$ $\mathfrak{Q}_{\text{\ \ }2}^{22},$ $%
\mathfrak{Q}_{\text{ \ }1}^{20},$ $\mathfrak{Q}_{\text{ \ }0}^{21},$ $%
\mathfrak{Q}_{\text{ \ }1}^{22},$ $\mathfrak{Q}_{\text{\ \ }2}^{21},$ $%
\mathfrak{Q}_{\text{ \ \ \ }1}^{2-1},$ $\mathfrak{Q}_{\text{\ \ }2}^{20},$ $%
\mathfrak{Q}_{\text{ \ }0}^{22},$ $\mathfrak{Q}_{-2}^{21},$ $\mathfrak{Q}%
_{-1}^{22},$ and $\mathfrak{Q}_{\text{ \ \ \ }2}^{2-2}$ correspond to the
case when the total spin in the incoming and outgoing channels are equal
and. We obtained the analytical expressions for $\mathfrak{Q}_{\text{\ \ }%
v^{^{\prime }}}^{2v}$ independent partial amplitudes and corresponding
formula are given in Appendix A by Eqs. (\ref{A6A5}) - (\ref{A18M5}). It
should be mentioned that 13 independent amplitudes are listed as the
elements of $5\times 5$ diagonal matrix in the upper corner of the matrix (%
\ref{integermatrix}).

There are 30 amplitudes $\mathfrak{M}_{2\upsilon ^{\prime }\text{ \ }%
}^{1\upsilon }$and $\mathfrak{M}_{1\upsilon ^{\prime }\text{ \ }}^{2\upsilon
}$ for the spin-mixing states for the collision of particles with the spins $%
1/2$ and $3/2.$ The number of these amplitudes can be reduced to 14 due to
parity conservation and considering that 2 spin-mixing partial amplitudes
equal 0. We obtained the analytical expressions for 7 independent $\mathfrak{%
M}_{2\upsilon ^{\prime }\text{ \ }}^{1\upsilon }$and 7 independent $%
\mathfrak{M}_{1\upsilon ^{\prime }\text{ \ }}^{2\upsilon }$ partial
spin-mixing amplitudes. The results of calculations \ of the corresponding
amplitudes are given in Appendix A3 by Eqs. (\ref{A31}) - (\ref{A32}).

\section{\protect\bigskip Processes with spin-$1/2$ $-$ spin-$1$}

\label{Spin1}

In this case according to the (\ref{channelspin}) channel spin can be $S=1/2$
and $S=3/2$. For the channel with half-integer spin values $1/2$ and $3/2$
the $M-$matrix is represented as

\begin{eqnarray}
&&%
\begin{tabular}{lllllll}
$\ S\text{ }\upsilon $ & \ \ \ $\ \ \ \ \ \ \ \frac{3}{2}\frac{3}{2}$ & $\ \
\ \ \ \ \ \ \ \ \ \ \frac{3}{2}\frac{1}{2}$ & \ $\ \ \ \ \ \ \ \frac{3}{2}-%
\frac{1}{2}$ & $\ \ \ \ \ \ \ \ \ \frac{3}{2}-\frac{3}{2}$ & $\ \ \ \ \ \ \
\ \ \ \frac{1}{2}\frac{1}{2}$ & $\ \ \ \ \ \ \ \ \ \ \frac{1}{2}-\frac{1}{2}$
\\
$\ S^{\prime }\upsilon ^{\prime }$ &  &  &  &  &  &
\end{tabular}
\notag \\
M &=&%
\begin{tabular}{ll}
$\ \ \frac{3}{2}\frac{3}{2}$ &  \\
$\ \ \frac{3}{2}\frac{1}{2}$ &  \\
$\frac{3}{2}-\frac{1}{2}$ &  \\
$\frac{3}{2}-\frac{3}{2}$ &  \\
$\ \ \frac{1}{2}\frac{1}{2}$ &  \\
$\frac{1}{2}-\frac{1}{2}$ &
\end{tabular}%
\left[
\begin{array}{cccccc}
\mathcal{Q}_{\text{ \ \ \ \ \ }3/2}^{3/2\text{ }3/2} & \mathcal{Q}_{\text{ \
\ \ \ \ }3/2}^{3/2\text{ }1/2} & \mathcal{Q}_{\text{ \ \ \ \ \ \ }%
3/2}^{3/2-1/2} & -\mathcal{Q}_{\text{ \ \ \ }-3/2}^{3/2\text{ }3/2} &
\mathfrak{M}_{3/2\text{ }3/2}^{1/2\text{ }1/2} & \mathfrak{M}_{3/2\text{ }%
~3/2}^{1/2-1/2} \\
\mathcal{Q}_{\text{ \ \ \ \ }1/2}^{3/2\text{ }3/2} & \mathcal{Q}_{\text{ \ \
\ \ \ }1/2}^{3/2\text{ }1/2} & \mathcal{Q}_{\text{ \ \ \ \ \ }1/2}^{3/2-1/2}
& \mathcal{Q}_{\text{ \ \ \ }-1/2}^{3/2\text{ }3/2} & \mathfrak{M}_{3/2\text{
}1/2}^{1/2\text{ }1/2} & \mathfrak{M}_{3/2\text{ \ }1/2}^{1/2-1/2} \\
\mathcal{Q}_{\text{ \ \ \ }-1/2}^{3/2\text{ }3/2} & -\mathcal{Q}_{\text{ \ \
\ \ \ \ }1/2}^{3/2-1/2} & \mathcal{Q}_{\text{ \ \ \ \ \ }1/2}^{3/2\text{ }%
1/2} & -\mathcal{Q}_{\text{ \ \ \ \ }1/2}^{3/2\text{ }3/2} & \mathfrak{M}%
_{3/2\text{ \ }1/2}^{1/2-1/2} & -\mathfrak{M}_{3/2\text{ }1/2}^{1/2\text{ }%
1/2} \\
\mathcal{Q}_{\text{ \ \ \ }-3/2}^{3/2\text{ }3/2} & \mathcal{Q}_{\text{ \ \
\ \ \ \ }3/2}^{3/2-1/2} & -\mathcal{Q}_{\text{ \ \ \ \ \ }3/2}^{3/2\text{ }%
1/2} & \mathcal{Q}_{\text{ \ \ \ \ \ }3/2}^{3/2\text{ }3/2} & -\mathfrak{M}%
_{3/2\text{ }~3/2}^{1/2-1/2} & \mathfrak{M}_{3/2\text{ }3/2}^{1/2\text{ }1/2}
\\
\mathfrak{M}_{1/2\text{\ }1/2}^{3/2\text{ }3/2} & \mathfrak{M}_{1/2\text{\ }%
1/2}^{3/2\text{ }1/2} & \mathfrak{M}_{1/2\text{\ \ }1/2}^{3/2-1/2} &
\mathfrak{M}_{1/2\text{\ \ }1/2}^{3/2-3/2} & \mathcal{D}_{\text{ \ \ \ \ \ }%
1/2}^{1/2\text{ }1/2} & \mathcal{D}_{\text{ \ \ \ \ \ \ }1/2}^{1/2-1/2} \\
-\mathfrak{M}_{1/2\text{\ \ }1/2}^{3/2-3/2} & \mathfrak{M}_{1/2\text{\ \ }%
1/2}^{3/2-1/2} & -\mathfrak{M}_{1/2\text{\ }1/2}^{3/2\text{ }1/2} &
\mathfrak{M}_{1/2\text{\ }1/2}^{3/2\text{ }3/2} & -\mathcal{D}_{\text{ \ \ \
\ \ \ }1/2}^{1/2-1/2} & \mathcal{D}_{\text{ \ \ \ \ \ }1/2}^{1/2\text{ }1/2}%
\end{array}%
\right]  \label{half_integer_Matrix1}
\end{eqnarray}%
In matrix (\ref{half_integer_Matrix1}) as above we use as the Fraktur letter
$\mathfrak{M}$ for spin-mixing states, while the calligraphic letters $%
\mathcal{D}$ and $\mathcal{Q}$ have been denoted for the doublet and quintet
spin states. The matrix (\ref{half_integer_Matrix1}) is defined by 36
partial amplitudes. However, according to (\ref{Parity}) the number of
partial amplitudes of this $M-$matrix can be reduced from 36 to 18
independent partial amplitudes.

The differential cross section for elastic scattering in a system of two
particles with spin-$1/2-$ spin-$1$, taking into account the spin-orbit
interaction, is presented in the form:

\begin{equation}
\frac{d\sigma \left( \theta \right) }{d\Omega }=\frac{1}{3}\frac{d\sigma _{%
\mathcal{D}}}{d\Omega }+\frac{2}{3}\frac{d\sigma _{\mathcal{Q}}}{d\Omega }.
\label{cross_sec2}
\end{equation}%
The channel spin $S$ can take values $1/2$ (doublet) and $3/2$ (quartet). In
this case according to Table \ref{tab1} $n$, $p$, $^{3}$H(1/2$^{+}$), $^{3}$%
He(1/2$^{+}$) and $^{2}$H(1$^{+}$), $^{6}$Li(1$^{+}$), $^{10}$Li(1$^{-}$), $%
^{12}$B(1$^{+}$) can be considered as colliding particles.

The doublet state, which corresponds to the channel spin $S=S^{\prime }=1/2$%
, is described by two independent amplitudes $\mathcal{D}_{\text{ \ \ \ \ \ }%
1/2}^{1/2\text{ }1/2}$ and $\mathcal{D}_{\text{ \ \ \ \ \ \ }1/2}^{1/2-1/2}$
of the matrix (\ref{half_integer_Matrix1}), which are presented by 2$\times $%
2 matrix in (\ref{half_integer_Matrix1}). The differential cross section for
the elastic scattering in the $doublet$ channel is well known and can be
written as

\begin{equation}
\frac{d\sigma _{\mathcal{D}}}{d\Omega }=\left\vert \mathcal{D}_{\text{ \ \ \
\ \ }1/2}^{1/2\text{ }1/2}\right\vert ^{2}+\left\vert \mathcal{D}_{\text{ \
\ \ \ \ \ }1/2}^{1/2-1/2}\right\vert ^{2}.  \label{doublet}
\end{equation}%
The cross section for the doublet spin channel is defined by two independent
partial amplitudes. In Eq. (\ref{doublet}) the expressions for the
independent partial amplitudes $\mathcal{D}_{\text{ \ \ \ \ \ }1/2}^{1/2%
\text{ }1/2}$ and $\mathcal{D}_{\text{ \ \ \ \ \ \ }1/2}^{1/2-1/2}$ are
given by Eqs. (\ref{B1A2}) - (\ref{B2B2}) in Appendix B.

The description of the $quartet$ spin channel, $S=S^{\prime }=3/2$, there
are 16 partial amplitudes. However, as it follows from (\ref{Parity}), there
are only 8 independent $\mathcal{Q}_{\text{ \ \ \ \ \ }\upsilon ^{^{\prime
}}}^{3/2\text{ }\upsilon }$ amplitudes of the matrix (\ref%
{half_integer_Matrix1}) you need to describe the quartet spin channel. The
differential cross section for the elastic scattering for a quartet state is
defined as

\begin{eqnarray}
\frac{d\sigma _{\mathcal{Q}}}{d\Omega } &=&\frac{1}{2}\left( \left\vert
\mathcal{Q}_{\text{ \ \ \ \ \ }1/2}^{3/2\text{ }1/2}\right\vert
^{2}+\left\vert \mathcal{Q}_{\text{ \ \ \ \ \ }3/2}^{3/2\text{ }%
3/2}\right\vert ^{2}+\left\vert \mathcal{Q}_{\text{ \ \ \ \ \ \ }%
1/2}^{3/2-1/2}\right\vert ^{2}+\left\vert \mathcal{Q}_{\text{ \ \ \ \ \ }%
3/2}^{3/2\text{ }1/2}\right\vert ^{2}+\left\vert \mathcal{Q}_{\text{ \ \ \ \
\ \ }3/2}^{3/2-1/2}\right\vert ^{2}\right.  \notag \\
&&\left. +\left\vert \mathcal{Q}_{\text{ \ \ \ \ }1/2}^{3/2\text{ }%
3/2}\right\vert ^{2}+\left\vert \mathcal{Q}_{\text{ \ \ \ }-1/2}^{3/2\text{ }%
3/2}\right\vert ^{2}+\left\vert \mathcal{Q}_{\text{ \ \ \ }-3/2}^{3/2\text{ }%
3/2}\right\vert ^{2}\right) .  \label{quartet}
\end{eqnarray}%
The analytical expressions for the independent partial quartet amplitudes $%
\mathcal{Q}_{\text{ \ \ \ \ \ }1/2}^{3/2\text{ }1/2},$ $\mathcal{Q}_{\text{
\ \ \ \ \ }3/2}^{3/2\text{ }3/2},$ $\mathcal{Q}_{\text{ \ \ \ \ \ \ }%
1/2}^{3/2-1/2},$ $\mathcal{Q}_{\text{ \ \ \ \ \ }3/2}^{3/2\text{ }1/2},$ $%
\mathcal{Q}_{\text{ \ \ \ \ \ \ }3/2}^{3/2-1/2},$ $\mathcal{Q}_{\text{ \ \ \
\ }1/2}^{3/2\text{ }3/2},$ $\mathcal{Q}_{\text{ \ \ \ }-1/2}^{3/2\text{ }%
3/2},$ and $\mathcal{Q}_{\text{ \ \ \ }-3/2}^{3/2\text{ }3/2}$ in Eq. (\ref%
{half_integer_Matrix1}) are calculated and presented in Appendix B by Eqs. (%
\ref{B3C4}) - (\ref{B10J4}).

There are 16 partial amplitudes that determine the spin-mixing states.
According to Eq. (\ref{Parity}) this number can be reduced to 8. We obtained
the analytical expression for the spin-mixing independent partial $\mathfrak{%
M}_{3/2\text{ }\upsilon ^{^{\prime }}}^{1/2\text{ }\upsilon }$and $\mathfrak{%
M}_{1/2\text{ }\upsilon ^{^{\prime }}}^{3/2\text{ }\upsilon }$ amplitudes of
the $M-$matrix for the half-integer channel spin $S=1/2$ and $S=3/2$. The
corresponding amplitudes are presented in Appendix B.3 by Eqs. (\ref{B21}) -
(\ref{B22}).

\section{\protect\bigskip Processes with spin-$1/2$ $-$ spin-$2$ and spin-$1$
$-$ spin-$3/2$}

\label{Spin2}

For the channel with the spin value $S=S^{\prime }=5/2$ of the incoming and
outgoing channel the part of $M-$matrix corresponding to the sextet spin
channel is represented as

\begin{eqnarray}
&&%
\begin{tabular}{lllllll}
$\ S\text{ }\upsilon $ & $\ \ \ \ \ \ \ \ \ \frac{5}{2}\frac{5}{2}$ & $\ \ \
\ \ \ \ \ \ \frac{5}{2}\frac{3}{2}$ & $\ \ \ \ \ \ \ \ \frac{5}{2}\frac{1}{2}
$ & $\ \ \ \ \ \ \ \ \frac{5}{2}-\frac{1}{2}$ & $\ \ \ \ \ \ \frac{5}{2}-%
\frac{3}{2}$ & $\ \ \ \ \ \frac{5}{2}-\frac{5}{2}$ \\
$\ S\upsilon ^{\prime }$ &  &  &  &  &  &
\end{tabular}
\notag \\
M &=&%
\begin{tabular}{ll}
$\ \ \frac{5}{2}\frac{5}{2}$ &  \\
$\ \ \frac{5}{2}\frac{3}{2}$ &  \\
$\ \ \frac{5}{2}\frac{1}{2}$ &  \\
$\frac{5}{2}-\frac{1}{2}$ &  \\
$\frac{5}{2}-\frac{3}{2}$ &  \\
$\frac{5}{2}-\frac{5}{2}$ &
\end{tabular}%
\left[
\begin{array}{cccccc}
\mathcal{S}_{\text{ \ \ \ \ \ }5/2}^{5/2\text{ }5/2} & \mathcal{S}_{\text{ \
\ \ \ \ }5/2}^{5/2\text{ }3/2} & \mathcal{S}_{\text{ \ \ \ \ \ }5/2}^{5/2%
\text{ }1/2} & -\mathcal{S}_{\text{ \ \ \ }-5/2}^{5/2\text{ }1/2} & \mathcal{%
S}_{\text{ \ \ \ \ \ \ }5/2}^{5/2-3/2} & -\mathcal{S}_{\text{ \ \ \ }%
-5/2}^{5/2\text{ }5/2} \\
\mathcal{S}_{\text{ \ \ \ \ \ }3/2}^{5/2\text{ }5/2} & \mathcal{S}_{\text{ \
\ \ \ \ }3/2}^{5/2\text{ }3/2} & \mathcal{S}_{\text{ \ \ \ \ \ }3/2}^{5/2%
\text{ }1/2} & \mathcal{S}_{\text{ \ \ \ \ \ \ }3/2}^{5/2-1/2} & -\mathcal{S}%
_{\text{ \ \ \ }-3/2}^{5/2\text{ }3/2} & \mathcal{S}_{\text{ \ \ \ \ \ \ }%
3/2}^{5/2-5/2} \\
\mathcal{S}_{\text{ \ \ \ \ \ }1/2}^{5/2\text{ }5/2} & \mathcal{S}_{\text{ \
\ \ \ \ }1/2}^{5/2\text{ }3/2} & \mathcal{S}_{\text{ \ \ \ \ \ }1/2}^{5/2%
\text{ }1/2} & -\mathcal{S}_{\text{ \ \ \ }-1/2}^{5/2\text{ }1/2} & \mathcal{%
S}_{\text{ \ \ \ \ \ \ }1/2}^{5/2-3/2} & -\mathcal{S}_{\text{ \ \ \ }%
-1/2}^{5/2\text{ }5/2} \\
\mathcal{S}_{\text{ \ \ \ }-1/2}^{5/2\text{ }5/2} & \mathcal{S}_{\text{ \ \
\ \ \ \ }1/2}^{5/2-3/2} & \mathcal{S}_{\text{ \ \ \ }-1/2}^{5/2\text{ }1/2}
& \mathcal{S}_{\text{ \ \ \ \ \ }1/2}^{5/2\text{ }1/2} & -\mathcal{S}_{\text{
\ \ \ \ \ }1/2}^{5/2\text{ }3/2} & \mathcal{S}_{\text{ \ \ \ \ \ }1/2}^{5/2%
\text{ }5/2} \\
\mathcal{S}_{\text{ \ \ \ \ \ \ }3/2}^{5/2-5/2} & \mathcal{S}_{\text{ \ \ \ }%
-3/2}^{5/2\text{ }3/2} & \mathcal{S}_{\text{ \ \ \ \ \ \ }3/2}^{5/2-1/2} & -%
\mathcal{S}_{\text{ \ \ \ \ \ }3/2}^{5/2\text{ }1/2} & \mathcal{S}_{\text{ \
\ \ \ \ }3/2}^{5/2\text{ }3/2} & -\mathcal{S}_{\text{ \ \ \ \ \ }3/2}^{5/2%
\text{ }5/2} \\
\mathcal{S}_{\text{ \ \ \ }-5/2}^{5/2\text{ }5/2} & \mathcal{S}_{\text{ \ \
\ \ \ \ }5/2}^{5/2-3/2} & \mathcal{S}_{\text{ \ \ \ }-5/2}^{5/2\text{ }1/2}
& \mathcal{S}_{\text{ \ \ \ \ \ }5/2}^{5/2\text{ }1/2} & -\mathcal{S}_{\text{
\ \ \ \ \ }5/2}^{5/2\text{ }3/2} & \mathcal{S}_{\text{ \ \ \ \ \ }5/2}^{5/2%
\text{ }5/2}%
\end{array}%
\right]  \label{half_int_matrix2}
\end{eqnarray}

In the interaction of particles with the spins $1/2$ and $2$, the channel
spin can take the values $S=3/2$ (quartet state, $d\sigma _{\mathcal{Q}%
}/d\Omega $) and $S=5/2$ (sextet state, $d\sigma _{\mathcal{S}}/d\Omega $).
In this case the following particles from Table \ref{tab1} which give the
quartet and sextet channel spin can be involved in the collision process: $n$%
, $p$, $^{3}$H(1/2$^{+}$), $^{3}$He(1/2$^{+}$) and $^{4}$H(2$^{-}$), $^{3}$%
Li(2$^{+}$), $^{4}$Li(2$^{-}$) $^{8}$B(2$^{+}$), $^{14}$B(2$^{-}$), $^{10}$%
N(2$^{-}$), $^{16}$N(2$^{-}$). The corresponding differential cross section
for the elastic scattering is

\begin{equation}
\frac{d\sigma \left( \theta \right) }{d\Omega }=\frac{2}{5}\frac{d\sigma _{%
\mathcal{Q}}}{d\Omega }+\frac{3}{5}\frac{d\sigma _{\mathcal{S}}}{d\Omega }.
\label{cross_sec3}
\end{equation}%
The partial differential cross section for the channel's spin $S=5/2$ is
represented by the following combination of amplitudes of the matrix (\ref%
{half_int_matrix2})

\begin{eqnarray}
\frac{d\sigma _{\mathcal{S}}}{d\Omega } &=&\frac{1}{2}\left[ \left\vert
\mathcal{S}_{\text{ \ \ \ \ \ }1/2}^{5/2\text{ }1/2}\right\vert
^{2}+\left\vert \mathcal{S}_{\text{ \ \ \ \ \ }3/2}^{5/2\text{ }%
3/2}\right\vert \right. +\left\vert \mathcal{S}_{\text{ \ \ \ \ \ }5/2}^{5/2%
\text{ }5/2}\right\vert ^{2}+\left\vert \mathcal{S}_{\text{ \ \ \ }-1/2}^{5/2%
\text{ }1/2}\right\vert ^{2}+\left\vert \mathcal{S}_{\text{ \ \ \ \ \ }%
3/2}^{5/2\text{ }1/2}\right\vert ^{2}+\left\vert \mathcal{S}_{\text{ \ \ \ \
\ }1/2}^{5/2\text{ }3/2}\right\vert ^{2}  \notag \\
&&+\left\vert \mathcal{S}_{\text{ \ \ \ \ \ }5/2}^{5/2\text{ }%
3/2}\right\vert ^{2}+\left\vert \mathcal{S}_{\text{ \ \ \ \ \ }3/2}^{5/2%
\text{ }5/2}\right\vert ^{2}+\left\vert \mathcal{S}_{\text{ \ \ \ \ \ \ }%
3/2}^{5/2-1/2}\right\vert ^{2}+\left\vert \mathcal{S}_{\text{ \ \ \ \ \ }%
5/2}^{5/2\text{ }1/2}\right\vert ^{2}+\left\vert \mathcal{S}_{\text{ \ \ \ \
\ \ }1/2}^{5/2-3/2}\right\vert ^{2}+\left\vert \mathcal{S}_{\text{ \ \ \ \ \
}1/2}^{5/2\text{ }5/2}\right\vert ^{2}  \notag \\
&&+\left\vert \mathcal{S}_{\text{ \ \ \ }-5/2}^{5/2\text{ }1/2}\right\vert
^{2}+\left\vert \mathcal{S}_{\text{ \ \ \ }-3/2}^{5/2\text{ }3/2}\right\vert
^{2}+\left\vert \mathcal{S}_{\text{ \ \ \ }-1/2}^{5/2\text{ }5/2}\right\vert
^{2}+\left\vert \mathcal{S}_{\text{ \ \ \ \ \ \ }5/2}^{5/2-3/2}\right\vert
^{2}+\left\vert \mathcal{S}_{\text{ \ \ \ \ \ \ }3/2}^{5/2-5/2}\right\vert
^{2}+\left. \left\vert \mathcal{S}_{\text{ \ \ \ }-5/2}^{5/2\text{ }%
5/2}\right\vert ^{2}\right] ,
\end{eqnarray}%
where $\mathcal{S}_{\text{ \ \ \ \ \ }\upsilon ^{^{\prime }}}^{5/2\text{ }%
\upsilon }$ are 18 independent partial amplitudes and the corresponding
analytical expressions for $\mathcal{S}_{\text{ \ \ \ \ \ }1/2}^{5/2\text{ }%
1/2},$ $\mathcal{S}_{\text{ \ \ \ \ \ }3/2}^{5/2\text{ }3/2},$ $\mathcal{S}_{%
\text{ \ \ \ \ \ }5/2}^{5/2\text{ }5/2},$ $\mathcal{S}_{\text{ \ \ \ }%
-1/2}^{5/2\text{ }1/2},$ $\mathcal{S}_{\text{ \ \ \ \ \ }3/2}^{5/2\text{ }%
1/2},$ $\mathcal{S}_{\text{ \ \ \ \ \ }5/2}^{5/2\text{ }3/2},$ $\mathcal{S}_{%
\text{ \ \ \ \ \ }3/2}^{5/2\text{ }5/2},$ $\mathcal{S}_{\text{ \ \ \ \ \ \ }%
3/2}^{5/2-1/2},$ $\mathcal{S}_{\text{ \ \ \ \ \ }5/2}^{5/2\text{ }1/2},$ $%
\mathcal{S}_{\text{ \ \ \ \ \ \ }1/2}^{5/2-3/2},$ $\mathcal{S}_{\text{ \ \ \
\ \ }1/2}^{5/2\text{ }5/2},$ $\mathcal{S}_{\text{ \ \ \ }-5/2}^{5/2\text{ }%
1/2},$ $\mathcal{S}_{\text{ \ \ \ }-3/2}^{5/2\text{ }3/2},$ $\mathcal{S}_{%
\text{ \ \ \ }-1/2}^{5/2\text{ }5/2},$ $\mathcal{S}_{\text{ \ \ \ \ \ \ }%
5/2}^{5/2-3/2},$ $\mathcal{S}_{\text{ \ \ \ \ \ \ }3/2}^{5/2-5/2},$ and $%
\mathcal{S}_{\text{ \ \ \ }-5/2}^{5/2\text{ }5/2}$ amplitudes. Let us
emphasize that we labeled the sextet spin channel partial amplitudes by the
calligraphic letter $\mathcal{S}.$ Results of our calculations for 18
independent partial amplitudes $\mathcal{S}_{\text{ \ \ \ \ \ }v^{^{\prime
}}}^{5/2\text{ }v}$ are presented by Eqs. (\ref{C1A6}) - (\ref{C18R6}) in
Appendix C.

For the system with spin-$1/2-$spin-$2$ channel spin $S$ and $S^{\prime }$
can equals $1/2,$ $3/2$ and $5/2.$ In this case, the differential cross
section for elastic scattering is determined by the expression

\begin{equation}
\frac{d\sigma \left( \theta \right) }{d\Omega }=\frac{1}{6}\frac{d\sigma _{%
\mathcal{D}}}{d\Omega }+\frac{1}{3}\frac{d\sigma _{\mathcal{Q}}}{d\Omega }+%
\frac{1}{2}\frac{d\sigma _{\mathcal{S}}}{d\Omega }.  \label{cross_sec4}
\end{equation}%
The collision of particles $n$, $p$, $^{3}$H(1/2$^{+}$), $^{3}$He(1/2$^{+}$)
and $^{4}$H(2$^{-}$), $^{3}$Li(2$^{+}$), $^{4}$Li(2$^{-}$) $^{8}$B(2$^{+}$),
$^{14}$B(2$^{-}$), $^{10}$N(2$^{-}$), $^{16}$N(2$^{-}$) from Table \ref{tab1}%
, which leads to the channel spin $1/2,$ $3/2$ and $5/2,$ can be considered.
The corresponding independent partial amplitudes for $\frac{d\sigma _{%
\mathcal{D}}}{d\Omega },$ $\frac{d\sigma _{\mathcal{Q}}}{d\Omega },$ and $%
\frac{d\sigma _{\mathcal{S}}}{d\Omega }$ cross sections are given in
Appendixes B and C by Eqs. (\ref{C1A6}) - (\ref{C18R6}) and (\ref{C1A6}) - (%
\ref{C18R6}). The general form of the 12$\times $12 matrix for the processes
with spin-$1/2$ $-$ spin-$2$ and spin-$1$ $-$ spin-$3/2$ is presented in
Ref. \cite{Tkachenko}. The number of independent matrix elements can be
reduced from 144 to 72. The doublet spin state with the cannel spin $S=1/2$
is described by only 2 partial amplitudes, while the discription of the
quartet channel with spin $S=3/2$ and the sextet spin state with $S=5/2$ are
required 8 and 18 independent matrix elements, respectively. This matrix
also contains 8 independent spin-mixing amplitudes for the mixing doublet
and quartet states. The analytical expressions for all these independent
partial amplitudes are presented in Appendixes. However, in the general case
of the 12$\times $12 matrix there are also 12 independent spin-mixing
amplitudes $\mathfrak{M}_{5/2\upsilon ^{\prime }}^{1/2\upsilon }$ and $%
\mathfrak{M}_{1/2\upsilon ^{\prime }}^{5/2\upsilon }$ for the mixing doublet
and sextet states, and 24 independent spin-mixing amplitudes $\mathfrak{M}%
_{5/2\upsilon ^{\prime }}^{3/2\upsilon }$ and $\mathfrak{M}_{3/2\upsilon
^{\prime }}^{5/2\upsilon }$ for the mixing quartet and sextet states. The
spin-mixing amplitudes for the latter two cases can be obtained.

\section{\protect\bigskip \qquad Conclusions}

\label{Conclusions}

We present the analytical expressions for the differential cross sections
for elastic scattering of nuclear particles for channels with a spin value
of 1/2, 1, 3/2, 2 and 5/2. The corresponding independent partial amplitudes
for each channel spin are obtained. These expressions are presented for
arbitrary orbital angular momentum $l$ and taking into account spin-orbit
splitting.\emph{\ }To describe the triplet state with the channel spin $S=1$%
, 5 independent amplitudes $\mathcal{T}_{\text{ \ }v^{^{\prime }}}^{1v}$ are
required. To describe the quintet ($S=2$), the number of independent
amplitudes $\mathfrak{Q}_{\text{\ \ }v^{^{\prime }}}^{2v}$ increases to 13.
The number of independent spin-mixing amplitudes $\mathfrak{M}_{2~\upsilon
^{\prime }}^{1~\upsilon }$ and $\mathfrak{M}_{1~\upsilon ^{\prime
}}^{2~\upsilon }$ in this case is 14. For the description of the
half-integer doublet ($S=1/2$) channel spin state only 2 independent partial
amplitudes $\mathcal{D}_{\text{ \ \ \ \ }v^{\prime }}^{1/2\text{ }v}$ are
required, while for the quartet state with the channel spin $S=3/2$ the
number of independent amplitudes $\mathcal{Q}_{\text{ \ \ \ \ \ }\upsilon
^{^{\prime }}}^{3/2\text{ }\upsilon }$\ equals 8. There are 18 independent
partial amplitudes $\mathcal{S}_{\text{ \ \ \ \ \ }v^{^{\prime }}}^{5/2\text{
}v}$ for the sextet state with the channel spin $S=5/2$. In the case of
mixing doublet and quartet states 8 independent spin-mixing amplitudes $%
\mathfrak{M}_{3/2~\upsilon ^{\prime }}^{1/2~\upsilon }$ and $\mathfrak{M}%
_{1/2~\upsilon ^{\prime }}^{3/2~\upsilon }$ are required. Obviously, with an
increase of the channel spin, the number of required independent amplitudes
for a correct description of scattering processes increases. For low energy
processes, whose description requires a small number of partial waves,
general expressions for the partial amplitudes can be reduced to simple
algebraic expressions.

Using experimental data for a nucleon-nucleus and nuclear--nuclear reaction
cross sections the obtained expressions for the differential cross sections
and independent partial amplitudes allow perform the phase shifts analyses
for different integer and half-integer channel spins, and find corresponding
phase shifts. These phase shifts will be used in the most essential stages
of construction of a nucleon-nuclear or nucleon-nucleon potentials, which
will be further employed for calculations of main characteristics of nuclear
reactions, in particular, astrophysical reactions at low energies.

\section{Acknowledgement}

A. S. Tkachenko acknowledges the support by al-Farabi Kazakh National University, the Physics Department and the Center for Theoretical Physics of New York City College of Technology, CUNY.

\appendix\bigskip

\section{\protect\bigskip Partial amplitudes for the spin-$1-$spin-$3/2$
system}

\subsection{Triplet spin state}

The independent partial amplitudes that determine the cross section for the
scattering in triplet state are:

\begin{equation}
\mathcal{T}_{\text{ \ }0}^{10}=f_{c}(\theta )+\frac{1}{2ik}\sum\limits_{l=0}%
\left[ \left( l+1\right) \alpha ^{l+1}+l\alpha ^{l-1}\right] \exp \left(
2i\omega _{l}\right) P_{l}\left( \cos \theta \right) ,  \label{A1A3}
\end{equation}

\begin{equation}
\mathcal{T}_{\text{ \ }1}^{11}=f_{c}(\theta )+\frac{1}{4ik}\sum\limits_{l=0}%
\left[ \left( l+2\right) \alpha ^{l+1}+\left( 2l+1\right) \alpha ^{l}+\left(
l-1\right) \alpha ^{l-1}\right] \exp \left( 2i\omega _{l}\right) P_{l}\left(
\cos \theta \right) ,  \label{B3}
\end{equation}

\begin{equation}
\mathcal{T}_{\text{ \ }1}^{10}=\frac{1}{2\sqrt{2}ik}\sum\limits_{l=1}\left[
\alpha ^{l+1}-\alpha ^{l-1}\right] \exp \left( 2i\omega _{l}\right)
P_{l}^{1}\left( \cos \theta \right) ,  \label{C3}
\end{equation}

\begin{equation}
\mathcal{T}_{\text{\ \ }0}^{11}=\frac{1}{2\sqrt{2}ik}\sum\limits_{l=1}\frac{1%
}{l\left( l+1\right) }\left[ l\left( l+2\right) \alpha ^{l+1}-\left(
2l+1\right) \alpha ^{l}-\left( l-1\right) \left( l+1\right) \alpha ^{l-1}%
\right] \exp \left( 2i\omega _{l}\right) P_{l}^{1}\left( \cos \theta \right)
,  \label{D3}
\end{equation}

\begin{equation}
\mathcal{T}_{\text{ \ \ \ }1}^{1-1}=\frac{1}{4ik}\sum\limits_{l=2}\frac{1}{%
l\left( l+1\right) }\left[ \alpha ^{l+1}-\left( 2l+1\right) \alpha
^{l}+\left( l+1\right) \alpha ^{l-1}\right] \exp \left( 2i\omega _{l}\right)
P_{l}^{2}\left( \cos \theta \right) .  \label{A5E3}
\end{equation}

The value $\alpha ^{J}=\left( U_{sl}^{J^{\pi }}-1\right) $ for each state
with full momentum $J$ is entered here.


\subsection{Quintet spin state}

The independent partial amplitudes that determine the cross section for the
scattering in quintet state are:

\begin{eqnarray}
\mathfrak{Q}_{\text{ \ }0}^{20} &=&f_{c}(\theta )+\frac{1}{4ik}%
\sum\limits_{l=0}\left[ \frac{3\left( l+1\right) \left( l+2\right) }{2l+3}%
\alpha ^{l+2}+\frac{2l\left( l+1\right) \left( 2l+1\right) }{\left(
2l+3\right) \left( 2l-1\right) }\alpha ^{l}+\frac{3l\left( l-1\right) }{2l-1}%
\alpha ^{l-2}\right]  \notag \\
&&\times \exp \left( 2i\omega _{l}\right) P_{l}\left( \cos \theta \right) ,
\label{A6A5}
\end{eqnarray}

\begin{eqnarray}
\mathfrak{Q}_{\text{ \ }1}^{21} &=&f_{c}(\theta )+\frac{1}{4ik}%
\sum\limits_{l=0}\left[ \frac{2\left( l+1\right) \left( l+3\right) }{2l+3}%
\alpha ^{l+2}+l\alpha ^{l+1}+\frac{3\left( 2l+1\right) }{\left( 2l+3\right)
\left( 2l-1\right) }\alpha ^{l}+\left( l+1\right) \alpha ^{l-1}\right.
\notag \\
&&\left. +\frac{2l\left( l-2\right) }{2l-1}\alpha ^{l-2}\right] \exp \left(
2i\omega _{l}\right) P_{l}\left( \cos \theta \right) ,  \label{B5}
\end{eqnarray}

\begin{eqnarray}
\mathfrak{Q}_{\text{\ \ }2}^{22} &=&f_{c}(\theta )+\frac{1}{8ik}%
\sum\limits_{l=0}\left[ \frac{\left( l+3\right) \left( l+4\right) }{2l+3}%
\alpha ^{l+2}+2\left( l+3\right) \alpha ^{l+1}+\frac{6\left( l+2\right)
\left( 2l+1\right) \left( l-1\right) }{\left( 2l-1\right) \left( 2l+3\right)
}\alpha ^{l}+\right.  \notag \\
&&\left. 2\left( l-2\right) \alpha ^{l-1}+\frac{\left( l-2\right) \left(
l-3\right) }{2l-1}\alpha ^{l-2}\right] \exp \left( 2i\omega _{l}\right)
P_{l}\left( \cos \theta \right) ,  \label{C5}
\end{eqnarray}

\begin{eqnarray}
\mathfrak{Q}_{\text{ \ }1}^{20} &=&-\frac{\sqrt{6}}{4ik}\sum\limits_{l=2}%
\frac{1}{\left( 2l+3\right) \left( 2l-1\right) }\left[ \left( l+2\right)
\left( 2l-1\right) \alpha ^{l+2}-\left( 2l+1\right) \alpha ^{l}\right.
\notag \\
&&\left. -\left( 2l+3\right) \left( l-1\right) \alpha ^{l-2}\right] \exp
\left( 2i\omega _{l}\right) P_{l}^{2}\left( \cos \theta \right) ,  \label{D5}
\end{eqnarray}

\begin{eqnarray}
\mathfrak{Q}_{\text{ \ }0}^{21} &=&-\frac{\sqrt{6}}{4ik}\sum\limits_{l=1}%
\left[ \frac{l+3}{2l+3}\alpha ^{l+2}-\frac{1}{l+1}\alpha ^{l+1}-\frac{\left(
2l+1\right) \left( 3-l\left( l+1\right) \right) }{l\left( l+1\right) \left(
2l+3\right) \left( 2l-1\right) }\alpha ^{l}-\frac{1}{l}\alpha ^{l-1}\right.
\notag \\
&&\left. -\frac{l-2}{2l-1}\alpha ^{l-2}\right] \exp \left( 2i\omega
_{l}\right) P_{l}^{1}\left( \cos \theta \right) ,  \label{E5}
\end{eqnarray}

\begin{eqnarray}
\mathfrak{Q}_{\text{ \ }1}^{22} &=&-\frac{1}{4ik}\sum\limits_{l=1}\left[
\frac{\left( l+3\right) \left( l+4\right) }{\left( 2l+3\right) \left(
l+1\right) }\alpha ^{l+2}+\frac{\left( l+3\right) \left( l-2\right) }{%
l\left( l+1\right) }\alpha ^{l+1}-\frac{9\left( l+2\right) \left(
2l+1\right) \left( l-1\right) }{l\left( 2l-1\right) \left( 2l+3\right)
\left( l+1\right) }\alpha ^{l}\right.  \notag \\
&&\left. -\frac{\left( l+3\right) \left( l-2\right) }{l\left( l+1\right) }%
\alpha ^{l-1}-\frac{\left( l-2\right) \left( l-3\right) }{l\left(
2l-1\right) }\alpha ^{l-2}\right] \exp \left( 2i\omega _{l}\right)
P_{l}^{1}\left( \cos \theta \right) ,  \label{F5}
\end{eqnarray}

\begin{eqnarray}
\mathfrak{Q}_{\text{\ \ }2}^{21} &=&-\frac{1}{4ik}\sum\limits_{l=1}\left[
\frac{l+3}{2l+3}\alpha ^{l+2}+\alpha ^{l+1}-\frac{3\left( 2l+1\right) }{%
\left( 2l+3\right) \left( 2l-1\right) }\alpha ^{l}-\alpha ^{l-1}\right.
\notag \\
&&\left. -\frac{l-2}{2l-1}\alpha ^{l-2}\right] \exp \left( 2i\omega
_{l}\right) P_{l}^{1}\left( \cos \theta \right) ,  \label{G5}
\end{eqnarray}

\begin{eqnarray}
\mathfrak{Q}_{\text{ \ \ \ }1}^{2-1} &=&\frac{1}{4ik}\sum\limits_{l=2}\left[
\frac{2\left( l+3\right) }{\left( 2l+3\right) \left( l+2\right) }\alpha
^{l+2}-\frac{l+4}{\left( l+1\right) \left( l+2\right) }\alpha ^{l+1}-\frac{%
9\left( 2l+1\right) }{l\left( 2l-1\right) \left( 2l+3\right) \left(
l+1\right) }\alpha ^{l}\right.  \notag \\
&&\left. -\frac{l-3}{l\left( l-1\right) }\alpha ^{l-1}+\frac{l-2}{\left(
2l-1\right) \left( l-1\right) }\alpha ^{l-2}\right] \exp \left( 2i\omega
_{l}\right) P_{l}^{2}\left( \cos \theta \right) ,  \label{H5}
\end{eqnarray}

\begin{equation}
\mathfrak{Q}_{\text{\ \ }2}^{20}=-\frac{\sqrt{6}}{8ik}\sum\limits_{l=2}\frac{%
1}{\left( 2l+3\right) \left( l-1\right) }\left[ \left( 2l-1\right) \alpha
^{l+2}-2\left( 2l+1\right) \alpha ^{l}+\left( 2l+3\right) \alpha ^{l-2}%
\right] \exp \left( 2i\omega _{l}\right) P_{l}^{2}\left( \cos \theta \right)
\label{I5}
\end{equation}

\begin{eqnarray}
\mathfrak{Q}_{\text{ \ }0}^{22} &=&\frac{\sqrt{6}}{8ik}\sum\limits_{l=2}%
\left[ \frac{\left( l+3\right) \left( l+4\right) }{\left( 2l+3\right) \left(
l+1\right) \left( l+2\right) }\alpha ^{l+2}-\frac{4\left( l+3\right) }{%
l\left( l+1\right) \left( l+2\right) }\alpha ^{l+1}-\frac{4\left(
2l+1\right) \left( l+4\right) \left( l-1\right) }{l\left( l+1\right) \left(
2l+3\right) \left( 2l-1\right) }\alpha ^{l}\right.  \notag \\
&&\left. +\frac{4\left( l-2\right) }{l\left( l+1\right) \left( l-1\right) }%
\alpha ^{l-1}+\frac{\left( l-2\right) \left( l-3\right) }{l\left(
2l-1\right) \left( l-1\right) }\alpha ^{l-2}\right] \exp \left( 2i\omega
_{l}\right) P_{l}^{2}\left( \cos \theta \right) ,  \label{J5}
\end{eqnarray}

\begin{eqnarray}
\mathfrak{Q}_{-2}^{21} &=&-\frac{1}{4ik}\sum\limits_{l=3}\left[ \frac{1}{%
\left( 2l+3\right) \left( l+2\right) }\alpha ^{l+2}-\frac{1}{\left(
l+1\right) \left( l+2\right) }\alpha ^{l+1}-\frac{3\left( 2l+1\right) }{%
l\left( 2l-1\right) \left( 2l+3\right) \left( l+1\right) }\alpha ^{l}\right.
\notag \\
&&\left. +\frac{1}{l\left( l-1\right) }\alpha ^{l-1}-\frac{1}{\left(
2l-1\right) \left( l-1\right) }\alpha ^{l-2}\right] \exp \left( 2i\omega
_{l}\right) P_{l}^{3}\left( \cos \theta \right) ,  \label{K5}
\end{eqnarray}

\begin{eqnarray}
\mathfrak{Q}_{-1}^{22} &=&-\frac{1}{4ik}\sum\limits_{l=3}\left[ \frac{l+4}{%
\left( 2l+3\right) \left( l+1\right) \left( l+2\right) }\alpha ^{l+2}-\frac{%
l+6}{l\left( l+1\right) \left( l+2\right) }\alpha ^{l+1}+\frac{15\left(
2l+1\right) }{l\left( 2l-1\right) \left( 2l+3\right) \left( l+1\right) }%
\alpha ^{l}\right.  \notag \\
&&\left. +\frac{l-5}{l\left( l+1\right) \left( l-1\right) }\alpha ^{l-1}-%
\frac{l-3}{l\left( 2l-1\right) \left( l-1\right) }\alpha ^{l-2}\right] \exp
\left( 2i\omega _{l}\right) P_{l}^{3}\left( \cos \theta \right) ,  \label{L5}
\end{eqnarray}

\begin{eqnarray}
\mathfrak{Q}_{\text{ \ \ \ }2}^{2-2} &=&\frac{1}{8ik}\sum\limits_{l=4}\left[
\frac{1}{\left( 2l+3\right) \left( l+1\right) \left( l+2\right) }\alpha
^{l+2}-\frac{2}{l\left( l+1\right) \left( l+2\right) }\alpha ^{l+1}+\frac{%
6\left( 2l+1\right) }{l\left( 2l-1\right) \left( 2l+3\right) \left(
l+1\right) }\alpha ^{l}\right.  \notag \\
&&\left. -\frac{2}{l\left( l+1\right) \left( l-1\right) }\alpha ^{l-1}+\frac{%
1}{l\left( 2l-1\right) \left( l-1\right) }\alpha ^{l-2}\right] \exp \left(
2i\omega _{l}\right) P_{l}^{4}\left( \cos \theta \right) .  \label{A18M5}
\end{eqnarray}

\bigskip

\subsection{Spin-mixing $M-$matrix elements, integer spin channel values $%
S=1 $ and $S=2$}

\bigskip

\begin{eqnarray}
\mathfrak{M}_{22}^{11} &=&\frac{i}{2k}\sum\limits_{l}\left[ \frac{l+2}{\sqrt{%
8l\left( 2l+1\right) }}\exp \left( i\left( \omega _{l}+\omega _{l-1}\right)
\right) U_{2\left( l-1\right) 1l}^{l+1}P_{l-1}^{1}\right.  \notag \\
&&-\sqrt{\frac{l+2}{4l}}\exp \left( 2i\omega _{l}\right)
U_{2l1l}^{l+1}P_{l}^{1}-  \notag \\
&&+\sqrt{\frac{\left( l+2\right) \left( l+3\right) }{2\left( 2l+1\right)
\left( 2l+5\right) }}\exp \left( i\left( \omega _{l}+\omega _{l+1}\right)
\right) U_{2\left( l+1\right) 1l}^{l+1}P_{l+1}^{1}  \notag \\
&&-\sqrt{\frac{l+2}{4\left( l+3\right) }}\exp \left( i\left( \omega
_{l}+\omega _{l+2}\right) \right) U_{2\left( l+2\right) 1l}^{l+1}P_{l+2}^{1}
\notag \\
&&+\sqrt{\frac{\left( l+2\right) \left( l+1\right) }{8\left( l+3\right)
\left( 2l+5\right) }}\exp \left( i\left( \omega _{l}+\omega _{l+3}\right)
\right) U_{2\left( l+3\right) 1l}^{l+1}P_{l+3}^{1}  \notag \\
&&-\sqrt{\frac{\left( 2l+1\right) \left( l+1\right) }{8\left( l-1\right)
\left( 2l-1\right) }}\exp \left( i\left( \omega _{l}+\omega _{l-2}\right)
\right) U_{2\left( l-2\right) 1l}^{l}P_{l-2}^{1}  \notag \\
&&+\sqrt{\frac{2l+1}{4\left( l-1\right) }}\exp \left( i\left( \omega
_{l}+\omega _{l-1}\right) \right) U_{2\left( l-1\right) 1l}^{l}P_{l-1}^{1}
\notag \\
&&-\sqrt{\frac{2l+1}{4\left( 2l-1\right) \left( 2l+3\right) }}\exp \left(
2i\omega _{l}\right) U_{2l1l}^{l}P_{l}^{1}  \notag \\
&&+\sqrt{\frac{2l+1}{4\left( l+2\right) }}\exp \left( i\left( \omega
_{l}+\omega _{l+1}\right) \right) U_{2\left( l+1\right) 1l}^{l}P_{l+1}^{1}
\notag \\
&&-\sqrt{\frac{l\left( 2l+1\right) }{8\left( l+2\right) \left( 2l+3\right) }}%
\exp \left( i\left( \omega _{l}+\omega _{l+2}\right) \right) U_{2\left(
l+2\right) 1l}^{l}P_{l+2}^{1}  \notag \\
&&+\sqrt{\frac{l\left( l-1\right) }{8\left( 2l-3\right) \left( l-2\right) }}%
\exp \left( i\left( \omega _{l}+\omega _{l-3}\right) \right) U_{2\left(
l-3\right) 1l}^{l-1}P_{l-3}^{1}  \notag \\
&&-\sqrt{\frac{l-1}{4\left( l-2\right) }}\exp \left( i\left( \omega
_{l}+\omega _{l-2}\right) \right) U_{2\left( l-2\right) 1l}^{l-1}P_{l-2}^{1}
\notag \\
&&+\sqrt{\frac{\left( l-1\right) \left( 2l-1\right) }{4\left( 2l-3\right)
\left( 2l+1\right) }}\exp \left( i\left( \omega _{l}+\omega _{l-1}\right)
\right) U_{2\left( l-1\right) 1l}^{l-1}P_{l-1}^{1}  \notag \\
&&-\sqrt{\frac{l-1}{4\left( l+1\right) }}\exp \left( 2i\omega _{l}\right)
U_{2l1l}^{l-1}P_{l}^{1}  \notag \\
&&\left. +\frac{l-1}{\sqrt{8\left( l+1\right) \left( 2l+1\right) }}\exp
\left( i\left( \omega _{l}+\omega _{l+1}\right) \right) U_{2\left(
l+1\right) 1l}^{l-1}P_{l+1}^{1}\right] ,  \label{A31}
\end{eqnarray}

\bigskip

\bigskip
\begin{eqnarray}
\mathfrak{M}_{22}^{10} &=&\frac{i}{2k}\sum\limits_{l}\left[ \frac{1}{2\sqrt{%
l\left( 2l+1\right) }}\exp \left( i\left( \omega _{l}+\omega _{l-1^{\prime
}}\right) \right) U_{2\left( l-1\right) 1l}^{l+1}P_{l-1}^{2}\right.  \notag
\\
&&-\frac{1}{\sqrt{2l\left( l+2\right) }}\exp \left( 2i\omega _{l}\right)
U_{2l1l}^{l+1}P_{l}^{2}  \notag \\
&&+\sqrt{\frac{3\left( 2l+3\right) }{2\left( 2l+1\right) \left( 2l+5\right)
\left( l+2\right) }}\exp \left( i\left( \omega _{l}+\omega _{l+1}\right)
\right) U_{2\left( l+1\right) 1l}^{l+1}P_{l+1}^{2}  \notag \\
&&-\frac{1}{\sqrt{2\left( l+2\right) \left( l+3\right) }}\exp \left( i\left(
\omega _{l}+\omega _{l+2}\right) \right) U_{2\left( l+2\right)
1l}^{l+1}P_{l+2}^{2}  \notag \\
&&+\sqrt{\frac{l+1}{4\left( l+3\right) \left( l+2\right) \left( 2l+5\right) }%
}\exp \left( i\left( \omega _{l}+\omega _{l+3}\right) \right) U_{2\left(
l+3\right) 1l}^{l+1}P_{l+3}^{2}  \notag \\
&&-\sqrt{\frac{l}{4\left( 2l-3\right) \left( l-1\right) \left( l-2\right) }}%
\exp \left( i\left( \omega _{l}+\omega _{l-3}\right) \right) U_{2\left(
l-3\right) 1l}^{l-1}P_{l-3}^{2}  \notag \\
&&+\frac{1}{\sqrt{2\left( l-1\right) \left( l-2\right) }}\exp \left( i\left(
\omega _{l}+\omega _{l-2}\right) \right) U_{2\left( l-2\right)
1l}^{l-1}P_{l-2}^{2}  \notag \\
&&-\sqrt{\frac{3\left( 2l-1\right) }{2\left( 2l-3\right) \left( 2l+1\right)
\left( l-1\right) }}\exp \left( i\left( \omega _{l}+\omega _{l-1}\right)
\right) U_{2\left( l-1\right) 1l}^{l-1}P_{l-1}^{2}  \notag \\
&&+\frac{1}{\sqrt{2\left( l-1\right) \left( l+1\right) }}\exp \left(
2i\omega _{l}\right) U_{2l1l}^{l-1}P_{l}^{2}  \notag \\
&&\left. +\frac{1}{\sqrt{4\left( 2l+1\right) \left( l+1\right) }}\exp \left(
i\left( \omega _{l}+\omega _{l+1}\right) \right) U_{2\left( l+1\right)
1l}^{l-1}P_{l+1}^{2}\right] ,
\end{eqnarray}

\bigskip

\bigskip
\begin{eqnarray}
\mathfrak{M}_{2\text{ \ }2}^{1-1} &=&\frac{i}{2k}\sum\limits_{l}\left[ \sqrt{%
\frac{\left( l+2\right) \left( l-2\right) \left( l-3\right) }{8l\left(
2l+1\right) \left( l+1\right) }}\exp \left( i\left( \omega _{l}+\omega
_{l-1^{\prime }}\right) \right) U_{2\left( l-1\right)
1l}^{l+1}P_{l-1}^{1}\right.  \notag \\
&&+\frac{\sqrt{\left( l+2\right) \left( l+3\right) \left( l-2\right) \left(
l-1\right) }}{2l\left( l+1\right) }\exp \left( 2i\omega _{l}\right)
U_{2l1l}^{l+1}P_{l}^{1}  \notag \\
&&+\sqrt{\frac{3l\left( l+3\right) \left( l+4\right) \left( 2l+3\right)
\left( l-1\right) }{2\left( l+1\right) ^{2}\left( l+2\right) \left(
2l+1\right) \left( 2l+5\right) }}\exp \left( i\left( \omega _{l}+\omega
_{l+1}\right) \right) U_{2\left( l+1\right) 1l}^{l+1}P_{l+1}^{1}  \notag \\
&&-\sqrt{\frac{l\left( l+5\right) \left( l+4\right) }{4\left( l+1\right)
\left( l+2\right) \left( l+3\right) }}\exp \left( i\left( \omega _{l}+\omega
_{l+2}\right) \right) U_{2\left( l+2\right) 1l}^{l+1}P_{l+2}^{1}  \notag \\
&&-\sqrt{\frac{\left( l+5\right) \left( l+6\right) }{8\left( l+3\right)
\left( 2l+5\right) }}\exp \left( i\left( \omega _{l}+\omega _{l+3}\right)
\right) U_{2\left( l+3\right) 1l}^{l+1}P_{l+3}^{1}  \notag \\
&&-\sqrt{\frac{\left( 2l+1\right) \left( l-3\right) \left( l-4\right) }{%
8l\left( l-1\right) \left( 2l-1\right) }}\exp \left( i\left( \omega
_{l}+\omega _{l-2}\right) \right) U_{2\left( l-2\right) 1l}^{l}P_{l-2}^{1}
\notag \\
&&+\frac{\sqrt{\left( 2l+1\right) \left( l+2\right) \left( l-2\right) \left(
l-3\right) }}{2l\left( l-1\right) }\exp \left( i\left( \omega _{l}+\omega
_{l-1}\right) \right) U_{2\left( l-1\right) 1l}^{l}P_{l-1}^{1}  \notag \\
&&-\sqrt{\frac{2l+1}{2}}\exp \left( 2i\omega _{l}\right)
U_{2l1l}^{l}P_{l}^{1}  \notag \\
&&+\sqrt{\frac{\left( 2l+1\right) \left( l+3\right) \left( l+4\right) \left(
l-1\right) }{2l\left( l+1\right) ^{2}\left( l+2\right) }}\exp \left( i\left(
\omega _{l}+\omega _{l+1}\right) \right) U_{2\left( l+1\right)
1l}^{l}P_{l+1}^{1}  \notag \\
&&-\sqrt{\frac{\left( 2l+1\right) \left( l+4\right) \left( l+5\right) }{%
8\left( l+1\right) \left( l+2\right) \left( 2l+3\right) }}\exp \left(
i\left( \omega _{l}+\omega _{l+2}\right) \right) U_{2\left( l+2\right)
1l}^{l}P_{l+2}^{1}  \notag \\
&&+\sqrt{\frac{\left( l-4\right) \left( l-5\right) }{8\left( 2l-3\right)
\left( l-2\right) }}\exp \left( i\left( \omega _{l}+\omega _{l-3}\right)
\right) U_{2\left( l-3\right) 1l}^{l-1}P_{l-3}^{1}  \notag \\
&&+\sqrt{\frac{\left( l+1\right) \left( l-3\right) \left( l-4\right) }{%
4\left( l-2\right) ^{2}\left( l-1\right) }}\exp \left( i\left( \omega
_{l}+\omega _{l-2}\right) \right) U_{2\left( l-2\right) 1l}^{l-1}P_{l-2}^{1}
\notag \\
&&+\sqrt{\frac{3\left( l-1\right) \left( l-2\right) \left( l-3\right) \left(
2l-1\right) \left( l+1\right) \left( l+2\right) }{4\left( 2l-3\right) \left(
2l+1\right) }}\exp \left( i\left( \omega _{l}+\omega _{l-1}\right) \right)
U_{2\left( l-1\right) 1l}^{l-1}P_{l-1}^{1}  \notag \\
&&-\sqrt{\frac{\left( l+3\right) \left( l+2\right) \left( l-2\right) }{%
4l^{2}\left( l+1\right) }}\exp \left( 2i\omega _{l}\right)
U_{2l1l}^{l-1}P_{l}^{1}  \notag \\
&&\left. +\sqrt{\frac{\left( l+3\right) \left( l+4\right) \left( l-1\right)
}{8l\left( l+1\right) \left( 2l+1\right) }}\exp \left( i\left( \omega
_{l}+\omega _{l+1}\right) \right) U_{2\left( l+1\right) 1l}^{l-1}P_{l+1}^{1}%
\right] ,
\end{eqnarray}

\bigskip

\bigskip
\begin{eqnarray}
\mathfrak{M}_{21}^{11} &=&\frac{i}{2k}\sum\limits_{l}\left[ \frac{l+2}{\sqrt{%
2\left( 2l+1\right) \left( l-1\right) }}\exp \left( i\left( \omega
_{l}+\omega _{l-1^{\prime }}\right) \right) U_{2\left( l-1\right)
1l}^{l+1}P_{l-1}^{1}\right.  \notag \\
&&-\sqrt{\frac{\left( l+2\right) }{4\left( l+1\right) }}\exp \left( 2i\omega
_{l}\right) U_{2l1l}^{l+1}P_{l}^{1}  \notag \\
&&-\sqrt{\frac{3\left( 2l+3\right) }{4\left( l+1\right) \left( 2l+1\right)
\left( 2l+5\right) }}\exp \left( i\left( \omega _{l}+\omega _{l+1}\right)
\right) U_{2\left( l+1\right) 1l}^{l+1}P_{l+1}^{1}  \notag \\
&&+\frac{1}{2}\exp \left( i\left( \omega _{l}+\omega _{l+2}\right) \right)
U_{2\left( l+2\right) 1l}^{l+1}P_{l+2}^{1}  \notag \\
&&-\sqrt{\frac{\left( l+2\right) \left( l+1\right) }{2\left( l+4\right)
\left( 2l+5\right) }}\exp \left( i\left( \omega _{l}+\omega _{l+3}\right)
\right) U_{2\left( l+3\right) 1l}^{l+1}P_{l+3}^{1}  \notag \\
&&-\sqrt{\frac{\left( 2l+1\right) \left( l+1\right) }{2\left( l-2\right)
\left( 2l-1\right) }}\exp \left( i\left( \omega _{l}+\omega _{l-2}\right)
\right) U_{2\left( l-2\right) 1l}^{l}P_{l-2}^{1}  \notag \\
&&+\sqrt{\frac{2l+1}{4l}}\exp \left( i\left( \omega _{l}+\omega
_{l-1}\right) \right) U_{2\left( l-1\right) 1l}^{l}P_{l-1}^{1}  \notag \\
&&+\frac{\sqrt{3}\left( 2l+1\right) }{\sqrt{4l\left( 2l-1\right) \left(
2l+3\right) \left( l+1\right) }}\exp \left( 2i\omega _{l}\right)
U_{2l1l}^{l}P_{l}^{1}  \notag \\
&&-\sqrt{\frac{2l+1}{4\left( l+1\right) }}\exp \left( i\left( \omega
_{l}+\omega _{l+1}\right) \right) U_{2\left( l+1\right) 1l}^{l}P_{l+1}^{1}
\notag \\
&&+\sqrt{\frac{l\left( 2l+1\right) }{2\left( l+3\right) \left( 2l+3\right) }}%
\exp \left( i\left( \omega _{l}+\omega _{l+2}\right) \right) U_{2\left(
l+2\right) 1l}^{l}P_{l+2}^{1}  \notag \\
&&+\sqrt{\frac{l\left( l-1\right) }{2\left( 2l-3\right) \left( l-3\right) }}%
\exp \left( i\left( \omega _{l}+\omega _{l-3}\right) \right) U_{2\left(
l-3\right) 1l}^{l-1}P_{l-3}^{1}  \notag \\
&&-\frac{1}{2}\exp \left( i\left( \omega _{l}+\omega _{l-2}\right) \right)
U_{2\left( l-2\right) 1l}^{l-1}P_{l-2}^{1}  \notag \\
&&-\sqrt{\frac{3\left( 2l-1\right) }{4l\left( 2l-3\right) \left( 2l+1\right)
}}\exp \left( i\left( \omega _{l}+\omega _{l-1}\right) \right) U_{2\left(
l-1\right) 1l}^{l-1}P_{l-1}^{1}  \notag \\
&&+\sqrt{\frac{l-1}{4l}}\exp \left( 2i\omega _{l}\right)
U_{2l1l}^{l-1}P_{l}^{1}  \notag \\
&&\left. -\frac{l-1}{\sqrt{2\left( l+2\right) \left( 2l+1\right) }}\exp
\left( i\left( \omega _{l}+\omega _{l+1}\right) \right) U_{2\left(
l+1\right) 1l}^{l-1}P_{l+1}^{1}\right] ,
\end{eqnarray}

\bigskip
\begin{eqnarray}
\mathfrak{M}_{21}^{10} &=&\frac{i}{2k}\sum\limits_{l}\left[ \frac{l+1}{\sqrt{%
l\left( 2l+1\right) }}\exp \left( i\left( \omega _{l}+\omega _{l-1^{\prime
}}\right) \right) U_{2\left( l-1\right) 1l}^{l+1}P_{l-1}^{1}\right.  \notag
\\
&&-\sqrt{\frac{l+2}{2l}}\exp \left( 2i\omega _{l}\right)
U_{2l1l}^{l+1}P_{l}^{1}  \notag \\
&&+\sqrt{\frac{3\left( 2l+3\right) }{2\left( l+2\right) \left( 2l+1\right)
\left( 2l+5\right) }}\exp \left( i\left( \omega _{l}+\omega _{l+1}\right)
\right) U_{2\left( l+1\right) 1l}^{l+1}P_{l+1}^{1}  \notag \\
&&+\frac{l+1}{\sqrt{2\left( l+2\right) \left( l+3\right) }}\exp \left(
i\left( \omega _{l}+\omega _{l+2}\right) \right) U_{2\left( l+2\right)
1l}^{l+1}P_{l+2}^{1}  \notag \\
&&-\sqrt{\frac{\left( l+2\right) \left( l+1\right) }{\left( l+3\right)
\left( 2l+5\right) }}\exp \left( i\left( \omega _{l}+\omega _{l+3}\right)
\right) U_{2\left( l+3\right) 1l}^{l+1}P_{l+3}^{1}  \notag \\
&&-\sqrt{\frac{l\left( l-1\right) }{\left( 2l-3\right) \left( l-2\right) }}%
\exp \left( i\left( \omega _{l}+\omega _{l-3}\right) \right) U_{2\left(
l-3\right) 1l}^{l-1}P_{l-3}^{1}  \notag \\
&&+\frac{l}{\sqrt{2\left( l-1\right) \left( l-2\right) }}\exp \left( i\left(
\omega _{l}+\omega _{l-2}\right) \right) U_{2\left( l-2\right)
1l}^{l-1}P_{l-2}^{1}  \notag \\
&&-\sqrt{\frac{3\left( 2l-1\right) }{2\left( 2l-3\right) \left( 2l+1\right)
\left( l-1\right) }}\exp \left( i\left( \omega _{l}+\omega _{l-1}\right)
\right) U_{2\left( l-1\right) 1l}^{l-1}P_{l-1}^{1}  \notag \\
&&-\sqrt{\frac{l-1}{2\left( l+1\right) }}\exp \left( 2i\omega _{l}\right)
U_{2l1l}^{l-1}P_{l}^{1}  \notag \\
&&\left. -\frac{l}{\sqrt{\left( l+1\right) \left( 2l+1\right) }}\exp \left(
i\left( \omega _{l}+\omega _{l+1}\right) \right) U_{2\left( l+1\right)
1l}^{l-1}P_{l+1}^{1}\right] ,
\end{eqnarray}

\bigskip
\begin{eqnarray}
\mathfrak{M}_{2\text{ \ }1}^{1-1} &=&-\frac{i}{2k}\sum\limits_{l}\left[
\frac{\left( l+2\right) \sqrt{l-2}}{\sqrt{2l\left( 2l+1\right) \left(
l+1\right) }}\exp \left( i\left( \omega _{l}+\omega _{l-1^{\prime }}\right)
\right) U_{2\left( l-1\right) 1l}^{l+1}P_{l-1}^{1}\right.  \notag \\
&&+\frac{\left( l+4\right) \sqrt{l-1}}{4\left( l+1\right) \sqrt{l}}\exp
\left( 2i\omega _{l}\right) U_{2l1l}^{l+1}P_{l}^{1}  \notag \\
&&+3\sqrt{\frac{3l\left( l+3\right) \left( 2l+3\right) }{2\left( l+1\right)
^{2}\left( l+2\right) \left( 2l+1\right) \left( 2l+5\right) }}\exp \left(
i\left( \omega _{l}+\omega _{l+1}\right) \right) U_{2\left( l+1\right)
1l}^{l+1}P_{l+1}^{1}  \notag \\
&&-\frac{\left( l-1\right) \sqrt{l+4}}{\sqrt{2\left( l+1\right) \left(
l+2\right) \left( l+3\right) }}\exp \left( i\left( \omega _{l}+\omega
_{l+2}\right) \right) U_{2\left( l+2\right) 1l}^{l+1}P_{l+2}^{1}  \notag \\
&&-\sqrt{\frac{\left( l+1\right) \left( l+5\right) }{2\left( l+3\right)
\left( 2l+5\right) }}\exp \left( i\left( \omega _{l}+\omega _{l+3}\right)
\right) U_{2\left( l+3\right) 1l}^{l+1}P_{l+3}^{1}  \notag \\
&&-\sqrt{\frac{\left( 2l+1\right) \left( l+1\right) \left( l-3\right) }{%
2l\left( l-1\right) \left( 2l-1\right) }}\exp \left( i\left( \omega
_{l}+\omega _{l-2}\right) \right) U_{2\left( l-2\right) 1l}^{l}P_{l-2}^{1}
\notag \\
&&-\frac{\left( l+3\right) \sqrt{\left( 2l+1\right) \left( l-2\right) }}{2l%
\sqrt{\left( l-1\right) \left( l+1\right) }}\exp \left( i\left( \omega
_{l}+\omega _{l-1}\right) \right) U_{2\left( l-1\right) 1l}^{l}P_{l-1}^{1}
\notag \\
&&-\frac{3\left( 2l+1\right) \sqrt{3\left( l+2\right) \left( l-1\right) }}{%
l\left( l+1\right) \sqrt{2\left( 2l-1\right) \left( 2l+3\right) }}\exp
\left( 2i\omega _{l}\right) U_{2l1l}^{l}P_{l}^{1}  \notag \\
&&-\frac{\left( l-2\right) \sqrt{\left( 2l+1\right) \left( l+3\right) }}{%
2\left( l+1\right) \sqrt{l\left( l+2\right) }}\exp \left( i\left( \omega
_{l}+\omega _{l+1}\right) \right) U_{2\left( l+1\right) 1l}^{l}P_{l+1}^{1}
\notag \\
&&+\sqrt{\frac{l\left( l+4\right) \left( 2l+1\right) }{2\left( l+1\right)
\left( l+2\right) \left( 2l+3\right) }}\exp \left( i\left( \omega
_{l}+\omega _{l+2}\right) \right) U_{2\left( l+2\right) 1l}^{l}P_{l+2}^{1}
\notag \\
&&+\sqrt{\frac{l\left( l-4\right) }{2\left( 2l-3\right) \left( l-2\right) }}%
\exp \left( i\left( \omega _{l}+\omega _{l-3}\right) \right) U_{2\left(
l-3\right) 1l}^{l-1}P_{l-3}^{1}  \notag \\
&&+\frac{\left( l+2\right) \sqrt{l-3}}{2\sqrt{l\left( l-1\right) \left(
l-2\right) }}\exp \left( i\left( \omega _{l}+\omega _{l-2}\right) \right)
U_{2\left( l-2\right) 1l}^{l-1}P_{l-2}^{1}  \notag \\
&&+\frac{3\sqrt{3\left( l+1\right) \left( l-2\right) \left( 2l-1\right) }}{2l%
\sqrt{\left( l-1\right) \left( l-2\right) }}\exp \left( i\left( \omega
_{l}+\omega _{l-1}\right) \right) U_{2\left( l-1\right) 1l}^{l-1}P_{l-1}^{1}
\notag \\
&&-\frac{\left( l-3\right) \sqrt{\left( l+2\right) }}{2l\sqrt{\left(
l+1\right) }}\exp \left( 2i\omega _{l}\right) U_{2l1l}^{l-1}P_{l}^{1}  \notag
\\
&&\left. -\frac{\left( l-1\right) \sqrt{l+3}}{\sqrt{2l\left( l+1\right)
\left( 2l+1\right) }}\exp \left( i\left( \omega _{l}+\omega _{l+1}\right)
\right) U_{2\left( l+1\right) 1l}^{l-1}P_{l+1}^{1}\right] ,
\end{eqnarray}

\bigskip

\bigskip
\begin{eqnarray}
\mathfrak{M}_{20}^{11} &=&\frac{i}{2k}\sum\limits_{l}\left[ \frac{l+2}{\sqrt{%
4l\left( 2l+1\right) }}\exp \left( i\left( \omega _{l}+\omega _{l-1^{\prime
}}\right) \right) U_{2\left( l-1\right) 1l}^{l+1}P_{l-1}^{1}\right.  \notag
\\
&&+\sqrt{\frac{3\left( l+2\right) }{2l\left( l+1\right) ^{2}}}\exp \left(
2i\omega _{l}\right) U_{2l1l}^{l+1}P_{l}^{1}  \notag \\
&&+\frac{\sqrt{\left( l+2\right) \left( 2l+3\right) }\left( 3-\left(
l+1\right) \left( l+2\right) \right) }{\sqrt{2\left( 2l+1\right) \left(
2l+5\right) }}\exp \left( i\left( \omega _{l}+\omega _{l+1}\right) \right)
U_{2\left( l+1\right) 1l}^{l+1}P_{l+1}^{1}  \notag \\
&&-\frac{1}{\sqrt{2\left( l+2\right) \left( l+3\right) }}\exp \left( i\left(
\omega _{l}+\omega _{l+2}\right) \right) U_{2\left( l+2\right)
1l}^{l+1}P_{l+2}^{1}  \notag \\
&&+\sqrt{\frac{3\left( l+2\right) \left( l+1\right) }{4\left( l+3\right)
\left( 2l+5\right) }}\exp \left( i\left( \omega _{l}+\omega _{l+3}\right)
\right) U_{2\left( l+3\right) 1l}^{l+1}P_{l+3}^{1}  \notag \\
&&-\sqrt{\frac{\left( 2l+1\right) \left( l+1\right) }{4\left( 2l-1\right)
\left( l-1\right) }}\exp \left( i\left( \omega _{l}+\omega _{l-2}\right)
\right) U_{2\left( l-2\right) 1l}^{l}P_{l-2}^{1}  \notag \\
&&-\sqrt{\frac{3\left( 2l+1\right) }{2l^{2}\left( l-1\right) }}\exp \left(
i\left( \omega _{l}+\omega _{l-1}\right) \right) U_{2\left( l-1\right)
1l}^{l}P_{l-1}^{1}  \notag \\
&&-\frac{\left( 2l+1\right) \left( 3-l\left( l+1\right) \right) }{l\left(
l+1\right) \sqrt{2\left( 2l+3\right) \left( 2l-1\right) }}\exp \left(
2i\omega _{l}\right) U_{2l1l}^{l}P_{l}^{1}  \notag \\
&&-\sqrt{\frac{3\left( 2l+1\right) }{2\left( l+1\right) ^{2}\left(
l+2\right) }}\exp \left( i\left( \omega _{l}+\omega _{l+1}\right) \right)
U_{2\left( l+1\right) 1l}^{l}P_{l+1}^{1}  \notag \\
&&-\sqrt{\frac{3l\left( 2l+1\right) }{4l+2\left( 2l+3\right) }}\exp \left(
i\left( \omega _{l}+\omega _{l+2}\right) \right) U_{2\left( l+2\right)
1l}^{l}P_{l+2}^{1}  \notag \\
&&+\sqrt{\frac{l\left( l-1\right) }{4\left( l-2\right) \left( 2l-3\right) }}%
\exp \left( i\left( \omega _{l}+\omega _{l-3}\right) \right) U_{2\left(
l-3\right) 1l}^{l-1}P_{l-3}^{1}  \notag \\
&&+\sqrt{\frac{3}{2\left( l-1\right) \left( l-2\right) }}\exp \left( i\left(
\omega _{l}+\omega _{l-2}\right) \right) U_{2\left( l-2\right)
1l}^{l-1}P_{l-2}^{1}  \notag \\
&&+\frac{\sqrt{\left( 2l-1\right) }\left( 3-l\left( l-1\right) \right) }{l%
\sqrt{2\left( l-1\right) \left( 2l+1\right) \left( 2l-3\right) }}\exp \left(
i\left( \omega _{l}+\omega _{l-1}\right) \right) U_{2\left( l-1\right)
1l}^{l-1}P_{l-1}^{1}  \notag \\
&&+\sqrt{\frac{3\left( l-1\right) }{2l^{2}\left( l+1\right) }}\exp \left(
2i\omega _{l}\right) U_{2l1l}^{l-1}P_{l}^{1}  \notag \\
&&\left. +\frac{\sqrt{3}\left( l-1\right) }{2\sqrt{\left( 2l+1\right) \left(
l+1\right) }}\exp \left( i\left( \omega _{l}+\omega _{l+1}\right) \right)
U_{2\left( l+1\right) 1l}^{l-1}P_{l+1}^{1}\right] ,
\end{eqnarray}

\bigskip

\begin{eqnarray}
\mathfrak{M}_{11}^{22} &=&-\frac{i}{2k}\sum\limits_{l}\left[ \frac{\left(
l+3\right) \left( l+4\right) }{2\left( l+2\right) \sqrt{2\left( 2l+3\right)
\left( l+1\right) }}\exp \left( i\left( \omega _{l}+\omega _{l+1}\right)
\right) U_{1\left( l+1\right) 2l}^{l+2}P_{l+1}^{1}\right.  \notag \\
&&-\frac{\left( l+4\right) \sqrt{\left( l+1\right) \left( 2l+5\right) }}{%
2\left( l+2\right) \sqrt{2\left( 2l+3\right) \left( l+1\right) }}\exp \left(
i\left( \omega _{l}+\omega _{l+2}\right) \right) U_{1\left( l+2\right)
2l}^{l+2}P_{l+2}^{1}  \notag \\
&&+\sqrt{\frac{\left( l+1\right) \left( l+2\right) }{8\left( 2l+3\right)
\left( l+3\right) }}\exp \left( i\left( \omega _{l}+\omega _{l+3}\right)
\right) U_{1\left( l+3\right) 2l}^{l+2}P_{l+3}^{1}  \notag \\
&&-\frac{\left( l+3\right) \sqrt{l+2}}{\left( l+1\right) \sqrt{2l}}\exp
\left( 2i\omega _{l}\right) U_{1l2l}^{l+1}P_{l}^{1}  \notag \\
&&+\frac{\left( l+3\right) \sqrt{l\left( 2l+3\right) }}{4\left( l+1\right)
\left( l+2\right) }\exp \left( i\left( \omega _{l}+\omega _{l+1}\right)
\right) U_{1\left( l+1\right) 2l}^{l+1}P_{l+1}^{1}  \notag \\
&&-\frac{\sqrt{l\left( l+1\right) }}{2\left( l+2\right) }\exp \left( i\left(
\omega _{l}+\omega _{l+2}\right) \right) U_{1\left( l+2\right)
2l}^{l+1}P_{l+2}^{1}  \notag \\
&&+\sqrt{\frac{3\left( l+1\right) \left( l+2\right) \left( 2l+1\right) }{%
4l^{2}\left( 2l+3\right) \left( 2l-1\right) }}\exp \left( i\left( \omega
_{l}+\omega _{l-1}\right) \right) U_{1\left( l-1\right) 2l}^{l}P_{l-1}^{1}
\notag \\
&&-\frac{3\left( l+2\right) \left( l-1\right) \left( 2l+1\right) }{l\left(
l+1\right) \sqrt{\left( 2l+3\right) \left( 2l-1\right) }}\exp \left(
2i\omega _{l}\right) U_{1l2l}^{l}P_{l}^{1}  \notag \\
&&+\frac{\left( l-1\right) \sqrt{3l\left( 2l+1\right) }}{2\left( l+1\right)
\sqrt{\left( 2l+3\right) \left( 2l-1\right) }}\exp \left( i\left( \omega
_{l}+\omega _{l+1}\right) \right) U_{1\left( l+1\right) 2l}^{l}P_{l+1}^{1}
\notag \\
&&-\frac{\sqrt{l\left( l+1\right) }}{2\left( l-1\right) }\exp \left( i\left(
\omega _{l}+\omega _{l-2}\right) \right) U_{1\left( l-2\right)
2l}^{l-1}P_{l-2}^{1}  \notag \\
&&+\frac{\sqrt{\left( l+1\right) \left( l-2\right) \left( 2l-1\right) }}{%
2\left( l-1\right) \sqrt{l}}\exp \left( i\left( \omega _{l}+\omega
_{l-1}\right) \right) U_{1\left( l-1\right) 2l}^{l-1}P_{l-1}^{1}  \notag \\
&&-\frac{\left( l-2\right) \sqrt{l-1}}{2l\sqrt{l+1}}\exp \left( 2i\omega
_{l}\right) U_{1l2l}^{l-1}P_{l}^{1}  \notag \\
&&+\sqrt{\frac{l\left( l-1\right) }{8\left( l-2\right) \left( 2l-1\right) }}%
\exp \left( i\left( \omega _{l}+\omega _{l-3}\right) \right) U_{1\left(
l-3\right) 2l}^{l-2}P_{l-3}^{1}  \notag \\
&&-\frac{\left( l-3\right) \sqrt{l\left( 2l-3\right) }}{2\left( l-1\right)
\sqrt{\left( l-2\right) \left( 2l-1\right) }}\exp \left( i\left( \omega
_{l}+\omega _{l-2}\right) \right) U_{1\left( l-2\right) 2l}^{l-2}P_{l-2}^{1}
\notag \\
&&\left. +\frac{\left( l-3\right) \left( l-2\right) }{2\left( l-1\right)
\sqrt{2l\left( 2l-1\right) }}\exp \left( i\left( \omega _{l}+\omega
_{l-1}\right) \right) U_{1\left( l-1\right) 2l}^{l-2}P_{l-1}^{1}\right]
\end{eqnarray}

\bigskip

\begin{eqnarray}
\mathfrak{M}_{11}^{21} &=&-\frac{i}{2k}\sum\limits_{l}\left[ \frac{\left(
l+3\right) \sqrt{\left( l+1\right) }}{\sqrt{2\left( 2l+3\right) }}\exp
\left( i\left( \omega _{l}+\omega _{l+1}\right) \right) U_{1\left(
l+1\right) 2l}^{l+2}P_{l+1}^{1}\right.  \notag \\
&&-\sqrt{\frac{\left( l+1\right) \left( l+3\right) \left( 2l+5\right) }{%
2\left( 2l+3\right) }}\exp \left( i\left( \omega _{l}+\omega _{l+2}\right)
\right) U_{1\left( l+2\right) 2l}^{l+2}P_{l+2}^{1}  \notag \\
&&+\sqrt{\frac{\left( l+1\right) \left( l+2\right) \left( l+3\right) }{%
2\left( 2l+3\right) }}\exp \left( i\left( \omega _{l}+\omega _{l+3}\right)
\right) U_{1\left( l+3\right) 2l}^{l+2}P_{l+3}^{1}  \notag \\
&&-\frac{\sqrt{l\left( l+2\right) }}{2}\exp \left( 2i\omega _{l}\right)
U_{1l2l}^{l+1}P_{l}^{1}  \notag \\
&&+\frac{\sqrt{l\left( 2l+3\right) }}{2}\exp \left( i\left( \omega
_{l}+\omega _{l+1}\right) \right) U_{1\left( l+1\right) 2l}^{l+1}P_{l+1}^{1}
\notag \\
&&-\frac{\sqrt{l\left( l+1\right) }}{2}\exp \left( i\left( \omega
_{l}+\omega _{l+2}\right) \right) U_{1\left( l+2\right) 2l}^{l+1}P_{l+2}^{1}
\notag \\
&&-\sqrt{\frac{3\left( l+1\right) \left( 2l+1\right) }{4\left( 2l+3\right)
\left( 2l-1\right) }}\exp \left( i\left( \omega _{l}+\omega _{l-1}\right)
\right) U_{1\left( l-1\right) 2l}^{l}P_{l-1}^{1}  \notag \\
&&+\frac{\sqrt{3}\left( 2l+1\right) }{2\sqrt{\left( 2l+3\right) \left(
2l-1\right) }}\exp \left( 2i\omega _{l}\right) U_{1l2l}^{l}P_{l}^{1}  \notag
\\
&&-\sqrt{\frac{3l\left( 2l+1\right) }{4\left( 2l+3\right) \left( 2l-1\right)
}}\exp \left( i\left( \omega _{l}+\omega _{l+1}\right) \right) U_{1\left(
l+1\right) 2l}^{l}P_{l+1}^{1}  \notag \\
&&+\frac{\sqrt{l\left( l+1\right) }}{2}\exp \left( i\left( \omega
_{l}+\omega _{l-2}\right) \right) U_{1\left( l-2\right) 2l}^{l-1}P_{l-2}^{1}
\notag \\
&&-\frac{\sqrt{\left( l+1\right) \left( 2l-1\right) }}{2}\exp \left( i\left(
\omega _{l}+\omega _{l-1}\right) \right) U_{1\left( l-1\right)
2l}^{l-1}P_{l-1}^{1}  \notag \\
&&+\frac{\sqrt{\left( l-1\right) \left( l+1\right) }}{2}\exp \left( 2i\omega
_{l}\right) U_{1l2l}^{l-1}P_{l}^{1}  \notag \\
&&-\sqrt{\frac{l\left( l-2\right) \left( l-1\right) }{2\left( 2l-1\right) }}%
\exp \left( i\left( \omega _{l}+\omega _{l-3}\right) \right) U_{1\left(
l-3\right) 2l}^{l-2}P_{l-3}^{1}  \notag \\
&&+\frac{\sqrt{l\left( l-2\right) \left( 2l-3\right) }}{\sqrt{2\left(
2l-1\right) }}\exp \left( i\left( \omega _{l}+\omega _{l-2}\right) \right)
U_{1\left( l-2\right) 2l}^{l-2}P_{l-2}^{1}  \notag \\
&&\left. -\frac{\left( l-2\right) \sqrt{l}}{\sqrt{2\left( 2l-1\right) }}\exp
\left( i\left( \omega _{l}+\omega _{l-1}\right) \right) U_{1\left(
l-1\right) 2l}^{l-2}P_{l-1}^{1}\right] ,
\end{eqnarray}

\begin{eqnarray}
\mathfrak{M}_{11}^{20} &=&\frac{i}{2k}\sum\limits_{l}\left[ \sqrt{\frac{%
3\left( l+1\right) }{4\left( 2l+3\right) }}\exp \left( i\left( \omega
_{l}+\omega _{l+1}\right) \right) U_{1\left( l+1\right)
2l}^{l+2}P_{l+1}^{1}\right.  \notag \\
&&-\sqrt{\frac{3\left( l+1\right) \left( 2l+5\right) }{4\left( 2l+3\right)
\left( l+3\right) }}\exp \left( i\left( \omega _{l}+\omega _{l+2}\right)
\right) U_{1\left( l+2\right) 2l}^{l+2}P_{l+2}^{1}  \notag \\
&&+\sqrt{\frac{3\left( l+1\right) \left( l+2\right) }{4\left( l+3\right)
\left( 2l+3\right) }}\exp \left( i\left( \omega _{l}+\omega _{l+3}\right)
\right) U_{1\left( l+3\right) 2l}^{l+2}P_{l+3}^{1}  \notag \\
&&-\sqrt{\frac{\left( 2l+1\right) \left( l+1\right) }{2\left( 2l+3\right)
\left( 2l-1\right) }}\exp \left( i\left( \omega _{l}+\omega _{l-1}\right)
\right) U_{1\left( l-1\right) 2l}^{l}P_{l-1}^{1}  \notag \\
&&+\frac{2l+1}{\sqrt{\left( 2l+3\right) \left( 2l-1\right) }}\exp \left(
2i\omega _{l}\right) U_{1l2l}^{l}P_{l}^{1}  \notag \\
&&-\sqrt{\frac{l\left( 2l+1\right) }{\left( 2l+3\right) \left( 2l-1\right) }}%
\exp \left( i\left( \omega _{l}+\omega _{l+1}\right) \right) U_{1\left(
l+1\right) 2l}^{l}P_{l+1}^{1}  \notag \\
&&+\sqrt{\frac{3l\left( l-1\right) }{4\left( l-2\right) \left( 2l-1\right) }}%
\exp \left( i\left( \omega _{l}+\omega _{l-3}\right) \right) U_{1\left(
l-3\right) 2l}^{l-2}P_{l-3}^{1}  \notag \\
&&-\sqrt{\frac{3l\left( 2l-3\right) }{2\left( l-2\right) \left( 2l-1\right) }%
}\exp \left( i\left( \omega _{l}+\omega _{l-2}\right) \right) U_{1\left(
l-2\right) 2l}^{l-2}P_{l-2}^{1}  \notag \\
&&\left. +\sqrt{\frac{3l}{2\left( 2l-1\right) }}\exp \left( i\left( \omega
_{l}+\omega _{l-1}\right) \right) U_{1\left( l-1\right) 2l}^{l-2}P_{l-1}^{1}%
\right] ,
\end{eqnarray}

\bigskip

\begin{eqnarray}
\mathfrak{M}_{1\text{ \ }1}^{2-1} &=&\frac{i}{2k}\sum\limits_{l}\left[ \frac{%
\sqrt{l+1}}{\left( l+2\right) \sqrt{2\left( 2l+3\right) }}\exp \left(
i\left( \omega _{l}+\omega _{l+1}\right) \right) U_{1\left( l+1\right)
2l}^{l+2}P_{l+1}^{2}\right.  \notag \\
&&+\frac{\sqrt{\left( l+1\right) \left( 2l+5\right) }}{\left( l+2\right)
\sqrt{2\left( 2l+3\right) \left( l+3\right) }}\exp \left( i\left( \omega
_{l}+\omega _{l+2}\right) \right) U_{1\left( l+2\right) 2l}^{l+2}P_{l+2}^{2}
\notag \\
&&+\sqrt{\frac{l+1}{2\left( l+3\right) \left( l+2\right) \left( 2l+3\right) }%
}\exp \left( i\left( \omega _{l}+\omega _{l+3}\right) \right) U_{1\left(
l+3\right) 2l}^{l+2}P_{l+3}^{2}  \notag \\
&&-\frac{\sqrt{l}}{2\left( l+1\right) \sqrt{\left( l+2\right) }}\exp \left(
2i\omega _{l}\right) U_{1l2l}^{l+1}P_{l}^{2}  \notag \\
&&-\frac{\sqrt{l\left( 2l+3\right) }}{2\left( l+1\right) \left( l+2\right) }%
\exp \left( i\left( \omega _{l}+\omega _{l+1}\right) \right) U_{1\left(
l+1\right) 2l}^{l+1}P_{l+1}^{2}  \notag \\
&&-\frac{\sqrt{l}}{2\left( l+2\sqrt{l+1}\right) }\exp \left( i\left( \omega
_{l}+\omega _{l+2}\right) \right) U_{1\left( l+2\right) 2l}^{l+1}P_{l+2}^{2}
\notag \\
&&-\frac{\sqrt{3\left( 2l+1\right) }}{2l\sqrt{\left( 2l+3\right) \left(
2l-1\right) \left( l+1\right) }}\exp \left( i\left( \omega _{l}+\omega
_{l-1}\right) \right) U_{1\left( l-1\right) 2l}^{l}P_{l-1}^{2}  \notag \\
&&-\frac{\sqrt{3\left( 2l+1\right) }}{2l\sqrt{\left( 2l+3\right) \left(
2l-1\right) \left( l+1\right) }}\exp \left( 2i\omega _{l}\right)
U_{1l2l}^{l}P_{l}^{2}  \notag \\
&&-\frac{\sqrt{3\left( 2l+1\right) }}{2\left( l+1\right) \sqrt{l\left(
2l+3\right) \left( 2l-1\right) }}\exp \left( i\left( \omega _{l}+\omega
_{l+1}\right) \right) U_{1\left( l+1\right) 2l}^{l}P_{l+1}^{2}  \notag \\
&&+\frac{\sqrt{l+1}}{2\left( l-1\right) \sqrt{l}}\exp \left( i\left( \omega
_{l}+\omega _{l-2}\right) \right) U_{1\left( l-2\right) 2l}^{l-1}P_{l-2}^{2}
\notag \\
&&+\frac{\sqrt{\left( l+1\right) \left( 2l-1\right) }}{2l\left( l-1\right) }%
\exp \left( i\left( \omega _{l}+\omega _{l-1}\right) \right) U_{1\left(
l-1\right) 2l}^{l-1}P_{l-1}^{2}  \notag \\
&&+\frac{\sqrt{l+1}}{2l\sqrt{l-1}}\exp \left( 2i\omega _{l}\right)
U_{1l2l}^{l-1}P_{l}^{2}  \notag \\
&&-\sqrt{\frac{l}{2\left( 2l-1\right) \left( l-1\right) \left( l-2\right) }}%
\exp \left( i\left( \omega _{l}+\omega _{l-3}\right) \right) U_{1\left(
l-3\right) 2l}^{l-2}P_{l-3}^{2}  \notag \\
&&-\frac{\sqrt{l\left( 2l-3\right) }}{\left( l-1\right) \sqrt{2\left(
2l-1\right) \left( l-2\right) }}\exp \left( i\left( \omega _{l}+\omega
_{l-2}\right) \right) U_{1\left( l-2\right) 2l}^{l-2}P_{l-2}^{2}  \notag \\
&&\left. -\frac{\sqrt{l}}{\left( l-1\right) \sqrt{2\left( 2l-1\right) }}\exp
\left( i\left( \omega _{l}+\omega _{l-1}\right) \right) U_{1\left(
l-1\right) 2l}^{l-2}P_{l-1}^{2}\right] ,
\end{eqnarray}

\bigskip
\begin{eqnarray}
\mathfrak{M}_{1\text{ \ }1}^{2-2} &=&\frac{i}{2k}\sum\limits_{l}\left[ \frac{%
1}{2\left( l+2\right) \sqrt{2\left( 2l+3\right) \left( l+1\right) }}\exp
\left( i\left( \omega _{l}+\omega _{l+1}\right) \right) U_{1\left(
l+1\right) 2l}^{l+2}P_{l+1}^{3}\right.  \notag \\
&&+\frac{1}{2\left( l+2\right) \sqrt{\left( 2l+3\right) \left( l+1\right)
\left( l+3\right) }}\exp \left( i\left( \omega _{l}+\omega _{l+2}\right)
\right) U_{1\left( l+2\right) 2l}^{l+2}P_{l+2}^{3}  \notag \\
&&+\frac{1}{\sqrt{8\left( l+3\right) \left( l+2\right) \left( l+1\right)
\left( 2l+3\right) }}\exp \left( i\left( \omega _{l}+\omega _{l+3}\right)
\right) U_{1\left( l+3\right) 2l}^{l+2}P_{l+3}^{3}  \notag \\
&&-\frac{1}{2\left( l+1\right) \sqrt{l\left( l+2\right) }}\exp \left(
2i\omega _{l}\right) U_{1l2l}^{l+1}P_{l}^{3}  \notag \\
&&-\frac{\sqrt{\left( 2l+3\right) }}{2\left( l+1\right) \left( l+2\right)
\sqrt{l}}\exp \left( i\left( \omega _{l}+\omega _{l+1}\right) \right)
U_{1\left( l+1\right) 2l}^{l+1}P_{l+1}^{3}  \notag \\
&&-\frac{1}{2\left( l+2\right) \sqrt{l\left( l+1\right) }}\exp \left(
i\left( \omega _{l}+\omega _{l+2}\right) \right) U_{1\left( l+2\right)
2l}^{l+1}P_{l+2}^{3}  \notag \\
&&+\frac{\sqrt{3\left( 2l+1\right) }}{2l\sqrt{\left( 2l+3\right) \left(
2l-1\right) \left( l+1\right) }}\exp \left( i\left( \omega _{l}+\omega
_{l-1}\right) \right) U_{1\left( l-1\right) 2l}^{l}P_{l-1}^{3}  \notag \\
&&+\frac{\left( 2l+1\right) \sqrt{3}}{2l\left( l+1\right) \sqrt{\left(
2l+3\right) \left( 2l-1\right) }}\exp \left( 2i\omega _{l}\right)
U_{1l2l}^{l}P_{l}^{3}  \notag \\
&&+\frac{\sqrt{3\left( 2l+1\right) }}{2\left( l+1\right) \sqrt{l\left(
2l+3\right) \left( 2l-1\right) }}\exp \left( i\left( \omega _{l}+\omega
_{l+1}\right) \right) U_{1\left( l+1\right) 2l}^{l}P_{l+1}^{3}  \notag \\
&&-\frac{1}{2\left( l-1\right) \sqrt{l\left( l+1\right) }}\exp \left(
i\left( \omega _{l}+\omega _{l-2}\right) \right) U_{1\left( l-2\right)
2l}^{l-1}P_{l-2}^{3}  \notag \\
&&-\frac{\sqrt{\left( 2l-1\right) }}{2\left( l-1\right) l\left( l+1\right) }%
\exp \left( i\left( \omega _{l}+\omega _{l-1}\right) \right) U_{1\left(
l-1\right) 2l}^{l-1}P_{l-1}^{3}  \notag \\
&&+\frac{1}{2l\sqrt{\left( l-1\right) \left( l+1\right) }}\exp \left(
2i\omega _{l}\right) U_{1l2l}^{l-1}P_{l}^{3}  \notag \\
&&+\frac{1}{\sqrt{8\left( 2l-1\right) l\left( l-1\right) \left( l-2\right) }}%
\exp \left( i\left( \omega _{l}+\omega _{l-3}\right) \right) U_{1\left(
l-3\right) 2l}^{l-2}P_{l-3}^{3}  \notag \\
&&+\frac{\sqrt{\left( 2l-3\right) }}{2\left( l-1\right) \sqrt{l\left(
2l-1\right) \left( l-2\right) }}\exp \left( i\left( \omega _{l}+\omega
_{l-2}\right) \right) U_{1\left( l-2\right) 2l}^{l-2}P_{l-2}^{3}  \notag \\
&&\left. +\frac{1}{2\left( l-1\right) \sqrt{2l\left( 2l-1\right) }}\exp
\left( i\left( \omega _{l}+\omega _{l-1}\right) \right) U_{1\left(
l-1\right) 2l}^{l-2}P_{l-1}^{3}\right] ,
\end{eqnarray}

\begin{eqnarray}
\mathfrak{M}_{10}^{22} &=&-\frac{i}{2k}\sum\limits_{l}\left[ \frac{l+4}{%
2\left( l+2\right) \sqrt{\left( l+1\right) \left( 2l+3\right) }}\exp \left(
i\left( \omega _{l}+\omega _{l+1}\right) \right) U_{1\left( l+1\right)
2l}^{l+2}P_{l+1}^{2}\right.  \notag \\
&&+\frac{\sqrt{\left( 2l+5\right) }}{\left( l+2\right) \sqrt{\left(
2l+3\right) \left( l+3\right) \left( l+1\right) }}\exp \left( i\left( \omega
_{l}+\omega _{l+2}\right) \right) U_{1\left( l+2\right) 2l}^{l+2}P_{l+2}^{2}
\notag \\
&&+\sqrt{\frac{\left( l+1\right) \left( l+3\right) }{4\left( l+2\right)
\left( 2l+3\right) }}\exp \left( i\left( \omega _{l}+\omega _{l+3}\right)
\right) U_{1\left( l+3\right) 2l}^{l+2}P_{l+3}^{2}  \notag \\
&&-\frac{l+3}{\sqrt{2}l\left( l+2\right) \left( l-1\right) }\exp \left(
2i\omega _{l}\right) U_{1l2l}^{l+1}P_{l}^{2}  \notag \\
&&-\frac{\sqrt{2\left( 2l+3\right) }}{\left( l+1\right) \left( l+2\right)
\sqrt{l}}\exp \left( i\left( \omega _{l}+\omega _{l+1}\right) \right)
U_{1\left( l+1\right) 2l}^{l+1}P_{l+1}^{2}  \notag \\
&&-\frac{\sqrt{l}}{\left( l+2\right) \sqrt{2\left( l+1\right) }}\exp \left(
i\left( \omega _{l}+\omega _{l+2}\right) \right) U_{1\left( l+2\right)
2l}^{l+1}P_{l+2}^{2}  \notag \\
&&-\frac{\left( l+2\right) \sqrt{3\left( 2l+1\right) }}{l\sqrt{2\left(
2l+3\right) \left( 2l-1\right) \left( l-1\right) }}\exp \left( i\left(
\omega _{l}+\omega _{l-1}\right) \right) U_{1\left( l-1\right)
2l}^{l}P_{l-1}^{2}  \notag \\
&&+\frac{\sqrt{6}\left( 2l+1\right) }{l\left( l+1\right) \sqrt{\left(
2l+3\right) \left( 2l-1\right) }}\exp \left( 2i\omega _{l}\right)
U_{1l2l}^{l}P_{l}^{2}  \notag \\
&&-\frac{\left( l-1\right) \sqrt{3\left( 2l+1\right) }}{\left( l+1\right)
\sqrt{2l\left( 2l+3\right) \left( 2l-1\right) }}\exp \left( i\left( \omega
_{l}+\omega _{l+1}\right) \right) U_{1\left( l+1\right) 2l}^{l}P_{l+1}^{2}
\notag \\
&&-\frac{\sqrt{l+1}}{\left( l-1\right) \sqrt{2l}}\exp \left( i\left( \omega
_{l}+\omega _{l-2}\right) \right) U_{1\left( l-2\right) 2l}^{l-1}P_{l-2}^{2}
\notag \\
&&-\frac{\sqrt{2\left( 2l-1\right) }}{l\left( l-1\right) \sqrt{l+1}}\exp
\left( i\left( \omega _{l}+\omega _{l-1}\right) \right) U_{1\left(
l-1\right) 2l}^{l-1}P_{l-1}^{2}  \notag \\
&&+\frac{l-2}{l\sqrt{2\left( l+1\right) \left( l-1\right) }}\exp \left(
2i\omega _{l}\right) U_{1l2l}^{l-1}P_{l}^{2}  \notag \\
&&+\sqrt{\frac{l}{4\left( 2l-1\right) \left( l-1\right) \left( l-2\right)
\left( l-4\right) }}\exp \left( i\left( \omega _{l}+\omega _{l-3}\right)
\right) U_{1\left( l-3\right) 2l}^{l-2}P_{l-3}^{2}  \notag \\
&&+\frac{\sqrt{2l-3}}{\left( l-1\right) \sqrt{l\left( 2l-1\right) \left(
l-2\right) }}\exp \left( i\left( \omega _{l}+\omega _{l-2}\right) \right)
U_{1\left( l-2\right) 2l}^{l-2}P_{l-2}^{2}  \notag \\
&&\left. -\frac{l-3}{2l\left( l-1\right) \sqrt{2l-1}}\exp \left( i\left(
\omega _{l}+\omega _{l-1}\right) \right) U_{1\left( l-1\right)
2l}^{l-2}P_{l-1}^{2}\right] ,
\end{eqnarray}

\begin{eqnarray}
\mathfrak{M}_{10}^{21} &=&-\frac{i}{2k}\sum\limits_{l}\left[ \frac{\left(
l+3\right) \sqrt{l+1}}{\left( l+2\right) \sqrt{2l+3}}\exp \left( i\left(
\omega _{l}+\omega _{l+1}\right) \right) U_{1\left( l+1\right)
2l}^{l+2}P_{l+1}^{1}\right.  \notag \\
&&+\frac{\sqrt{\left( l+1\right) \left( 2l+5\right) }}{\left( l+2\right)
\sqrt{\left( l+3\right) \left( 2l+3\right) }}\exp \left( i\left( \omega
_{l}+\omega _{l+2}\right) \right) U_{1\left( l+2\right) 2l}^{l+2}P_{l+2}^{1}
\notag \\
&&-\sqrt{\frac{\left( l+1\right) \left( l+2\right) }{\left( l+3\right)
\left( 2l+3\right) }}\exp \left( i\left( \omega _{l}+\omega _{l+3}\right)
\right) U_{1\left( l+3\right) 2l}^{l+2}P_{l+3}^{1}  \notag \\
&&-\frac{\sqrt{l\left( l+2\right) }}{\sqrt{2}\left( l+1\right) }\exp \left(
2i\omega _{l}\right) U_{1l2l}^{l+1}P_{l}^{1}  \notag \\
&&-\frac{\sqrt{l\left( 2l+3\right) }}{\sqrt{2}\left( l+1\right) \left(
l+2\right) }\exp \left( i\left( \omega _{l}+\omega _{l+1}\right) \right)
U_{1\left( l+1\right) 2l}^{l+1}P_{l+1}^{1}  \notag \\
&&-\frac{\sqrt{l\left( l+1\right) }}{\sqrt{2}\left( l+2\right) }\exp \left(
i\left( \omega _{l}+\omega _{l+2}\right) \right) U_{1\left( l+2\right)
2l}^{l+1}P_{l+2}^{1}  \notag \\
&&-\sqrt{\frac{3\left( l+1\right) \left( 2l+1\right) }{2l^{2}\left(
2l+3\right) \left( 2l-1\right) }}\exp \left( i\left( \omega _{l}+\omega
_{l-1}\right) \right) U_{1\left( l-1\right) 2l}^{l}P_{l-1}^{1}  \notag \\
&&-\frac{\sqrt{3}\left( 2l+1\right) }{l\left( l+1\right) \sqrt{2\left(
2l+3\right) \left( 2l-1\right) }}\exp \left( 2i\omega _{l}\right)
U_{1l2l}^{l}P_{l}^{1}  \notag \\
&&+\frac{\sqrt{3l\left( 2l+1\right) }}{\left( l+1\right) \sqrt{2\left(
2l+3\right) \left( 2l-1\right) }}\exp \left( i\left( \omega _{l}+\omega
_{l+1}\right) \right) U_{1\left( l+1\right) 2l}^{l}P_{l+1}^{1}  \notag \\
&&+\frac{\sqrt{l\left( l+1\right) }}{\sqrt{2}\left( l-1\right) }\exp \left(
i\left( \omega _{l}+\omega _{l-2}\right) \right) U_{1\left( l-2\right)
2l}^{l-1}P_{l-2}^{1}  \notag \\
&&+\frac{\sqrt{\left( l+1\right) \left( 2l-1\right) }}{\sqrt{2}l\left(
l-1\right) }\exp \left( i\left( \omega _{l}+\omega _{l-1}\right) \right)
U_{1\left( l-1\right) 2l}^{l-1}P_{l-1}^{1}  \notag \\
&&-\frac{\sqrt{\left( l-1\right) \left( l+1\right) }}{\sqrt{2}l}\exp \left(
2i\omega _{l}\right) U_{1l2l}^{l-1}P_{l}^{1}  \notag \\
&&-\sqrt{\frac{l\left( l-1\right) }{\left( 2l-1\right) \left( l-2\right) }}%
\exp \left( i\left( \omega _{l}+\omega _{l-3}\right) \right) U_{1\left(
l-3\right) 2l}^{l-2}P_{l-3}^{1}  \notag \\
&&-\frac{\sqrt{l\left( 2l-3\right) }}{\left( l-1\right) \sqrt{\left(
2l-1\right) \left( l-2\right) }}\exp \left( i\left( \omega _{l}+\omega
_{l-2}\right) \right) U_{1\left( l-2\right) 2l}^{l-2}P_{l-2}^{1}  \notag \\
&&\left. +\frac{\left( l-2\right) \sqrt{l}}{\left( l-1\right) \sqrt{2\left(
2l-1\right) }}\exp \left( i\left( \omega _{l}+\omega _{l-1}\right) \right)
U_{1\left( l-1\right) 2l}^{l-2}P_{l-1}^{1}\right] .  \label{A32}
\end{eqnarray}

\section{Partial amplitudes for spin-$1-$spin-$1/2$ system}

Below are defined the partial amplitudes for the doublet and quartet spin
states.

\subsection{Doublet spin state}

The independent partial amplitudes for the doublet spin state are the
following:

\begin{equation}
\mathcal{D}_{\text{ \ \ \ \ \ }1/2}^{1/2\text{ }1/2}=f_{c}(\theta )+\frac{1}{%
2ik}\sum\limits_{l=0}\left[ \left( l+1\right) \alpha ^{l+1/2}+l\alpha
^{l-1/2}\right] \exp \left( 2i\omega _{l}\right) P_{l}\left( \cos \theta
\right) ,  \label{B1A2}
\end{equation}

\begin{equation}
\mathcal{D}_{\text{ \ \ \ \ \ \ }1/2}^{1/2-1/2}=-\frac{1}{2ik}%
\sum\limits_{l=1}\sqrt{l\left( l+1\right) }\left[ \alpha ^{l+1/2}-\alpha
^{l-1/2}\right] \exp \left( 2i\omega _{l}\right) P_{l}^{1}\left( \cos \theta
\right) .,  \label{B2B2}
\end{equation}

\subsection{Quartet spin states}

The independent partial amplitudes for the quartet spin state are the
following:

\begin{eqnarray}
\mathcal{Q}_{\text{ \ \ \ \ \ }1/2}^{3/2\text{ }1/2} &=&f_{c}(\theta )+\frac{%
1}{4ik}\sum\limits_{l=0}\left[ \frac{3\left( l+1\right) \left( l+2\right) }{%
2l+3}\alpha ^{l+3/2}+\frac{l\left( l+1\right) }{2l+3}\alpha ^{l+1/2}+\frac{%
l\left( l+1\right) }{2l-1}\alpha ^{l-1/2}\right.  \notag \\
&&\left. +\frac{3l\left( l-1\right) }{2l-1}\alpha ^{l-3/2}\right] \exp
\left( 2i\omega _{l}\right) P_{l}\left( \cos \theta \right) ,  \label{B3C4}
\end{eqnarray}

\begin{eqnarray}
\mathcal{Q}_{\text{ \ \ \ \ \ }3/2}^{3/2\text{ }3/2} &=&f_{c}(\theta )+\frac{%
1}{4ik}\sum\limits_{l=0}\left[ \frac{\left( l+2\right) \left( l+3\right) }{%
2l+3}\alpha ^{l+3/2}+\frac{3\left( l+1\right) \left( l+2\right) }{2l+3}%
\alpha ^{l+1/2}+\frac{3l\left( l-1\right) }{2l-1}\alpha ^{l-1/2}\right.
\notag \\
&&\left. +\frac{\left( l-1\right) \left( l-2\right) }{2l-1}\alpha ^{l-3/2}%
\right] \exp \left( 2i\omega _{l}\right) P_{l}\left( \cos \theta \right) ,
\label{D4}
\end{eqnarray}

\begin{eqnarray}
\mathcal{Q}_{\text{ \ \ \ \ \ \ }1/2}^{3/2-1/2} &=&-\frac{1}{4ik}%
\sum\limits_{l=1}\left[ \frac{3\left( l+2\right) }{2l+3}\alpha ^{l+3/2}-%
\frac{l+3}{2l+3}\alpha ^{l+1/2}+\frac{l-2}{2l-1}\alpha ^{l-1/2}-\frac{%
3\left( l-1\right) }{2l-1}\alpha ^{l-3/2}\right]  \notag \\
&&\times \exp \left( 2i\omega _{l}\right) P_{l}^{1}\left( \cos \theta
\right) ,  \label{E4}
\end{eqnarray}

\begin{eqnarray}
\mathcal{Q}_{\text{ \ \ \ \ \ }3/2}^{3/2\text{ }1/2} &=&-\frac{\sqrt{3}}{4ik}%
\sum\limits_{l=1}\left[ \frac{l+2}{2l+3}\alpha ^{l+3/2}+\frac{l+1}{2l+3}%
\alpha ^{l+1/2}-\frac{l}{2l-1}\alpha ^{l-1/2}-\frac{l-1}{2l-1}\alpha ^{l-3/2}%
\right]  \notag \\
&&\times \exp \left( 2i\omega _{l}\right) P_{l}^{1}\left( \cos \theta
\right) ,  \label{F4}
\end{eqnarray}

\begin{eqnarray}
\mathcal{Q}_{\text{ \ \ \ \ \ \ }3/2}^{3/2-1/2} &=&-\frac{\sqrt{3}}{4ik}%
\sum\limits_{l=1}\left[ \frac{\left( l+2\right) \left( l+3\right) }{\left(
2l+3\right) \left( l+1\right) }\alpha ^{l+3/2}+\frac{\left( l+2\right)
\left( l-3\right) }{l\left( 2l+3\right) }\alpha ^{l+1/2}-\frac{\left(
l+4\right) \left( l-1\right) }{\left( 2l-1\right) \left( l+1\right) }\alpha
^{l-1/2}\right.  \notag \\
&&\left. -\frac{\left( l-1\right) \left( l-2\right) }{l\left( 2l-1\right) }%
\alpha ^{l-3/2}\right] \exp \left( 2i\omega _{l}\right) P_{l}\left( \cos
\theta \right) ,  \label{G4}
\end{eqnarray}

\begin{eqnarray}
\mathcal{Q}_{\text{ \ \ \ \ }1/2}^{3/2\text{ }3/2} &=&\frac{\sqrt{3}}{4ik}%
\sum\limits_{l=2}\left[ \frac{1}{2l+3}\alpha ^{l+3/2}-\frac{1}{2l+3}\alpha
^{l+1/2}-\frac{1}{2l-1}\alpha ^{l-1/2}+\frac{1}{2l-1}\alpha ^{l-3/2}\right]
\notag \\
&&\times \exp \left( 2i\omega _{l}\right) P_{l}^{2}\left( \cos \theta
\right) ,  \label{H4}
\end{eqnarray}

\begin{eqnarray}
\mathcal{Q}_{\text{ \ \ \ }-1/2}^{3/2\text{ }3/2} &=&\frac{\sqrt{3}}{4ik}%
\sum\limits_{l=2}\left[ \frac{l+3}{\left( 2l+3\right) \left( l+1\right) }%
\alpha ^{l+3/2}-\frac{l+6}{l\left( 2l+3\right) }\alpha ^{l+1/2}-\frac{l-5}{%
\left( 2l-1\right) \left( l+1\right) }\alpha ^{l-1/2}\right.  \notag \\
&&\left. +\frac{l-2}{l\left( 2l-1\right) }\alpha ^{l-3/2}\right] \exp \left(
2i\omega _{l}\right) P_{l}^{2}\left( \cos \theta \right) ,  \label{I4}
\end{eqnarray}

\begin{eqnarray}
\mathcal{Q}_{\text{ \ \ \ }-3/2}^{3/2\text{ }3/2} &=&-\frac{1}{4ik}%
\sum\limits_{l=3}\left[ \frac{1}{\left( 2l+3\right) \left( l+1\right) }%
\alpha ^{l+3/2}-\frac{3}{l\left( 2l+3\right) }\alpha ^{l+1/2}+\frac{3}{%
\left( 2l-1\right) \left( l+1\right) }\alpha ^{l-1/2}\right.  \notag \\
&&\left. -\frac{1}{l\left( 2l-1\right) }\alpha ^{l-3/2}\right] \left(
2i\omega _{l}\right) P_{l}^{3}\left( \cos \theta \right) .  \label{B10J4}
\end{eqnarray}

\subsection{\protect\bigskip Spin-mixing $M-$matrix elements for
half-integer channel spin $S=1/2$ and $S=3/2$}

\begin{eqnarray}
\mathfrak{M}_{3/2\text{ }3/2}^{1/2\text{ }1/2} &=&\frac{i}{2k}\sum\limits_{l}%
\left[ \frac{l+1}{\sqrt{2l\left( 2l+1\right) }}\exp \left( i\left( \omega
_{l}+\omega _{l-1}\right) \right) U_{3/2\left( l-1\right)
1/2l}^{~l+1/2}P_{l-1}^{1}\right.  \notag \\
&&-\frac{\sqrt{3}\left( l+1\right) }{\sqrt{2l\left( 2l+3\right) }}\exp
\left( 2i\omega _{l}\right) U_{3/2l1/2l}^{~l+1/2}P_{l}^{1}  \notag \\
&&-\frac{\sqrt{3}\left( l+1\right) }{\sqrt{2\left( l+2\right) \left(
2l+3\right) }}\exp \left( i\left( \omega _{l}+\omega _{l+1}\right) \right)
U_{3/2\left( l+1\right) 1/2l}^{~l+1/2}P_{l+1}^{1}  \notag \\
&&-\frac{l+1}{\sqrt{2\left( l+2\right) \left( 2l+3\right) }}\exp \left(
i\left( \omega _{l}+\omega _{l+2}\right) \right) U_{3/2\left( l+2\right)
1/2l}^{~l+1/2}P_{l+2}^{1}  \notag \\
&&-\frac{l}{\sqrt{2\left( l-1\right) \left( 2l-1\right) }}\exp \left(
i\left( \omega _{l}+\omega _{l-2}\right) \right) U_{3/2\left( l-2\right)
1/2l}^{~l-1/2}P_{l-2}^{1}  \notag \\
&&+\frac{\sqrt{3}l}{\sqrt{2\left( l-1\right) \left( 2l+1\right) }}\exp
\left( i\left( \omega _{l}+\omega _{l-1}\right) \right) U_{3/2\left(
l-1\right) 1/2l}^{~l-1/2}P_{l-1}^{1}  \notag \\
&&-\frac{\sqrt{3}l}{\sqrt{2\left( l+1\right) \left( 2l-1\right) }}\exp
\left( 2i\omega _{l}\right) U_{3/2l1/2l}^{~l-1/2}P_{l}^{1}  \notag \\
&&\left. +\frac{l}{\sqrt{2\left( l+1\right) \left( 2l+1\right) }}\exp \left(
i\left( \omega _{l}+\omega _{l+1}\right) \right) U_{3/2\left( l+1\right)
1/2l}^{~l-1/2}P_{l+1}^{1}\right]  \label{B21}
\end{eqnarray}

\bigskip
\begin{eqnarray}
\mathfrak{M}_{3/2\text{ }~3/2}^{1/2-1/2} &=&\frac{i}{2k}\sum\limits_{l}\left[
\frac{1}{\sqrt{2l\left( 2l+1\right) }}\exp \left( i\left( \omega _{l}+\omega
_{l-1}\right) \right) U_{3/2\left( l-1\right)
1/2l}^{~l+1/2}P_{l-1}^{2}\right.  \notag \\
&&+\sqrt{\frac{3\left( l+1\right) }{2l\left( 2l+3\right) \left( l-1\right) }}%
\exp \left( 2i\omega _{l}\right) U_{3/2l1/2l}^{~l+1/2}P_{l}^{2}  \notag \\
&&+\sqrt{\frac{3}{\left( 2l+1\right) \left( l+2\right) }}\exp \left( i\left(
\omega _{l}+\omega _{l+1}\right) \right) U_{3/2\left( l+1\right)
1/2l}^{~l+1/2}P_{l+1}^{2}  \notag \\
&&+\frac{1}{\sqrt{2\left( l+2\right) \left( 2l-1\right) }}\exp \left(
i\left( \omega _{l}+\omega _{l+2}\right) \right) U_{3/2\left( l+2\right)
1/2l}^{~l+1/2}P_{l+2}^{2}  \notag \\
&&-\frac{1}{\sqrt{2\left( l-1\right) \left( 2l-1\right) }}\exp \left(
i\left( \omega _{l}+\omega _{l-2}\right) \right) U_{3/2\left( l-2\right)
1/2l}^{~l-1/2}P_{l-2}^{2}  \notag \\
&&-\sqrt{\frac{3l}{2\left( 2l+3\right) \left( l-1\right) \left( l-2\right) }}%
\exp \left( i\left( \omega _{l}+\omega _{l-1}\right) \right) U_{3/2\left(
l-1\right) 1/2l}^{~l-1/2}P_{l-1}^{2}  \notag \\
&&-\sqrt{\frac{3}{2\left( 2l-1\right) \left( l+1\right) }}\exp \left(
2i\omega _{l}\right) U_{3/2l1/2l}^{~l-1/2}P_{l}^{2}  \notag \\
&&-\left. \frac{1}{\sqrt{2\left( l+1\right) \left( 2l+1\right) }}\exp \left(
i\left( \omega _{l}+\omega _{l+1}\right) \right) U_{3/2\left( l+1\right)
1/2l}^{~l-1/2}P_{l+1}^{2}\right]
\end{eqnarray}

\begin{eqnarray}
\mathfrak{M}_{3/2\text{ }1/2}^{1/2\text{ }1/2} &=&-\frac{i}{2k}%
\sum\limits_{l}\left[ \frac{\sqrt{3l}\left( l+1\right) }{\sqrt{2\left(
2l+1\right) }}\exp \left( i\left( \omega _{l}+\omega _{l-1}\right) \right)
U_{3/2\left( l-1\right) 1/2l}^{~l+1/2}P_{l-1}^{0}\right.  \notag \\
&&-\frac{\left( l+1\right) \sqrt{l}}{\sqrt{2\left( 2l+3\right) }}\exp \left(
2i\omega _{l}\right) U_{3/2l1/2l}^{~l-1/2}P_{l}^{0}  \notag \\
&&-\frac{\left( l+1\right) \sqrt{l+2}}{\sqrt{2\left( 2l+1\right) }}\exp
\left( i\left( \omega _{l}+\omega _{l+1}\right) \right) U_{3/2\left(
l+1\right) 1/2l}^{~l+1/2}P_{l+1}^{0}  \notag \\
&&+\frac{\left( l+1\right) \sqrt{3\left( l+2\right) }}{\sqrt{2\left(
2l+3\right) }}\exp \left( i\left( \omega _{l}+\omega _{l+2}\right) \right)
U_{3/2\left( l+2\right) 1/2l}^{~l+1/2}P_{l+2}^{0}  \notag \\
&&-\frac{l\sqrt{3\left( l-1\right) }}{\sqrt{2\left( 2l-1\right) }}\exp
\left( i\left( \omega _{l}+\omega _{l-2}\right) \right) U_{3/2\left(
l-2\right) 1/2l}^{~l-1/2}P_{l-2}^{0}  \notag \\
&&+\frac{l\sqrt{l-1}}{\sqrt{2\left( 2l+1\right) }}\exp \left( i\left( \omega
_{l}+\omega _{l-1}\right) \right) U_{3/2\left( l-1\right)
1/2l}^{~l-1/2}P_{l-1}^{0}  \notag \\
&&+\frac{l\sqrt{l+1}}{\sqrt{2\left( 2l-1\right) }}\exp \left( 2i\omega
_{l}\right) U_{3/2l1/2l}^{~l-1/2}P_{l}^{0}  \notag \\
&&\left. -\frac{l\sqrt{3\left( l+1\right) }}{\sqrt{2\left( 2l+1\right) }}%
\exp \left( i\left( \omega _{l}+\omega _{l+1}\right) \right) U_{3/2\left(
l+1\right) 1/2l}^{~l-1/2}P_{l+1}^{0}\right]
\end{eqnarray}

\begin{eqnarray}
\mathfrak{M}_{3/2\text{ \ }1/2}^{1/2-1/2} &=&-\frac{i}{2k}\sum\limits_{l}%
\left[ \frac{\sqrt{3l\left( l-1\right) }\left( l+1\right) }{\sqrt{2\left(
2l+1\right) }}\exp \left( i\left( \omega _{l}+\omega _{l-1}\right) \right)
U_{3/2\left( l-1\right) 1/2l}^{~l+1/2}P_{l-1}^{0}\right.  \notag \\
&&+\frac{\left( l+1\right) \sqrt{l+3}}{\sqrt{2\left( 2l+3\right) }}\exp
\left( 2i\omega _{l}\right) U_{3/2l1/2l}^{~l+1/2}P_{l}^{0}  \notag \\
&&-\frac{\left( l-1\right) \sqrt{l+1}}{\sqrt{2\left( 2l+3\right) }}\exp
\left( i\left( \omega _{l}+\omega _{l+1}\right) \right) U_{3/2\left(
l+1\right) 1/2l}^{~l+1/2}P_{l+1}^{0}  \notag \\
&&-\frac{\left( l+1\right) \sqrt{3\left( l+3\right) }}{\sqrt{2\left(
2l+3\right) }}\exp \left( i\left( \omega _{l}+\omega _{l+2}\right) \right)
U_{3/2\left( l+2\right) 1/2l}^{~l+1/2}P_{l+2}^{0}  \notag \\
&&-\frac{l\sqrt{3\left( l-2\right) }}{\sqrt{2\left( 2l-1\right) }}\exp
\left( i\left( \omega _{l}+\omega _{l-2}\right) \right) U_{3/2\left(
l-2\right) 1/2l}^{~l-1/2}P_{l-2}^{0}  \notag \\
&&-\frac{\left( l+1\right) \sqrt{l}}{\sqrt{2\left( 2l+1\right) }}\exp \left(
i\left( \omega _{l}+\omega _{l-1}\right) \right) U_{3/2\left( l-1\right)
1/2l}^{~l-1/2}P_{l-1}^{0}  \notag \\
&&+\frac{\left( l-2\right) \sqrt{l}}{\sqrt{2\left( 2l-1\right) }}\exp \left(
2i\omega _{l}\right) U_{3/2\left( l-1\right) 1/2l}^{~l-1/2}P_{l-1}^{0}
\notag \\
&&+\left. \frac{l\sqrt{3\left( l+2\right) }}{\sqrt{2\left( 2l+1\right) }}%
\exp \left( i\left( \omega _{l}+\omega _{l+1}\right) \right) U_{3/2\left(
l+1\right) 1/2l}^{~l-1/2}P_{l+1}^{0}\right]
\end{eqnarray}

\bigskip

\begin{eqnarray}
\mathfrak{M}_{1/2\text{\ }1/2}^{3/2\text{ }3/2} &=&-\frac{i}{2k}%
\sum\limits_{l}\left[ \frac{l+3}{\sqrt{2\left( 2l+3\right) \left( l+1\right)
}}\exp \left( i\left( \omega _{l}+\omega _{l+1}\right) \right)
U_{1/2(l+1)3/2l}^{~l+3/2}P_{l+1}^{1}\right.  \notag \\
&&-\sqrt{\frac{l+1}{2\left( 2l+3\right) }}\exp \left( i\left( \omega
_{l}+\omega _{l+2}\right) \right) U_{1/2(l+2)3/2l}^{~l+3/2}P_{l+2}^{1}
\notag \\
&&-\frac{l+2}{\sqrt{2l\left( 2l+3\right) }}\exp \left( 2i\omega _{l}\right)
U_{1/2l3/2l}^{~l+1/2}P_{l}^{1}  \notag \\
&&+\sqrt{\frac{3l}{2\left( 2l+3\right) }}\exp \left( i\left( \omega
_{l}+\omega _{l+1}\right) \right) U_{1/2(l+1)3/2l}^{~l+1/2}P_{l+1}^{1}
\notag \\
&&+\sqrt{\frac{l+1}{2\left( 2l-1\right) }}\exp \left( i\left( \omega
_{l}+\omega _{l-1}\right) \right) U_{1/2\left( l-1\right)
3/2l}^{~l-1/2}P_{l-1}^{1}  \notag \\
&&-\frac{l-1}{\sqrt{2\left( l+1\right) \left( 2l-1\right) }}\exp \left(
2i\omega _{l}\right) U_{1/2l3/2l}^{~l-1/2}P_{l}^{1}  \notag \\
&&-\sqrt{\frac{l}{2\left( 2l-1\right) }}\exp \left( i\left( \omega
_{l}+\omega _{l-2}\right) \right) U_{1/2(l-2)3/2l}^{~l-3/2}P_{l-2}^{1}
\notag \\
&&\left. +\frac{l-2}{\sqrt{2l\left( 2l-1\right) }}\exp \left( i\left( \omega
_{l}+\omega _{l-1}\right) \right) U_{1/2\left( l-1\right)
3/2l}^{~l-3/2}P_{l}^{1}\right]
\end{eqnarray}

\begin{eqnarray}
\mathfrak{M}_{1/2\text{\ }1/2}^{3/2\text{ }1/2} &=&-\frac{i}{2k}%
\sum\limits_{l}\left[ \frac{\left( l+2\right) \sqrt{l+3}}{\sqrt{2\left(
2l+3\right) }}\exp \left( i\left( \omega _{l}+\omega _{l+1}\right) \right)
U_{1/2(l+1)3/2l}^{~l+3/2}P_{l+1}^{0}\right.  \notag \\
&&-\frac{\left( l+2\right) \sqrt{l+3}}{\sqrt{2\left( 2l+3\right) }}\exp
\left( i\left( \omega _{l}+\omega _{l+2}\right) \right)
U_{1/2(l+2)3/2l}^{~l+3/2}P_{l+2}^{0}  \notag \\
&&-\frac{\left( l+1\right) \sqrt{3\left( l+2\right) }}{\sqrt{2\left(
2l+3\right) }}\exp \left( 2i\omega _{l}\right)
U_{1/2(l)3/2l}^{~l+1/2}P_{l}^{0}  \notag \\
&&+\frac{\left( l+1\right) \sqrt{3\left( l+2\right) }}{\sqrt{2\left(
2l+3\right) }}\exp \left( i\left( \omega _{l}+\omega _{l+1}\right) \right)
U_{1/2(l+1)3/2l}^{~l+1/2}P_{l+1}^{0}  \notag \\
&&+\frac{l\sqrt{3\left( l-1\right) }}{\sqrt{2\left( 2l-1\right) }}\exp
\left( i\left( \omega _{l}+\omega _{l-1}\right) \right)
U_{1/2(l-1)3/2l}^{~l-1/2}P_{l-1}^{0}  \notag \\
&&+\frac{l\sqrt{3\left( l-1\right) }}{\sqrt{2\left( 2l-1\right) }}\exp
\left( 2i\omega _{l}\right) U_{1/2l)3/2l}^{~l-1/2}P_{l}^{0}  \notag \\
&&-\frac{\left( l-1\right) \sqrt{\left( l-2\right) }}{\sqrt{2\left(
2l-1\right) }}\exp \left( i\left( \omega _{l}+\omega _{l-2}\right) \right)
U_{1/2(l-2)3/2l}^{~l-3/2}P_{l-2}^{0}  \notag \\
&&+\left. \frac{\left( l-1\right) \sqrt{\left( l-2\right) }}{\sqrt{2\left(
2l-1\right) }}\exp \left( i\left( \omega _{l}+\omega _{l-1}\right) \right)
U_{1/2(l-1)3/2l}^{~l-3/2}P_{l-1}^{0}\right]
\end{eqnarray}

\begin{eqnarray}
\mathfrak{M}_{1/2\text{\ \ }1/2}^{3/2-1/2} &=&-\frac{i}{2k}\sum\limits_{l}%
\left[ \sqrt{\frac{l+3}{2\left( 2l+3\right) }}\exp \left( i\left( \omega
_{l}+\omega _{l+1}\right) \right) U_{1/2(l+1)3/2l}^{~l+3/2}P_{l+1}^{1}\right.
\notag \\
&&+\sqrt{\frac{l+3}{2\left( 2l+3\right) }}\exp \left( i\left( \omega
_{l}+\omega _{l+2}\right) \right) U_{1/2(l+2)3/2l}^{~l+3/2}P_{l+2}^{1}
\notag \\
&&-\sqrt{\frac{3\left( l+2\right) }{2\left( 2l+3\right) }}\exp \left(
2i\omega _{l}\right) U_{1/2(l)3/2l}^{~l+1/2}P_{l}^{1}  \notag \\
&&-\sqrt{\frac{3\left( l+2\right) }{2\left( 2l+3\right) }}\exp \left(
i\left( \omega _{l}+\omega _{l+1}\right) \right)
U_{1/2(l+1)3/2l}^{~l+1/2}P_{l+1}^{1}  \notag \\
&&+\frac{\sqrt{3\left( l-1\right) }}{\sqrt{2\left( 2l-1\right) }}\exp \left(
i\left( \omega _{l}+\omega _{l-1}\right) \right)
U_{1/2(l-1)3/2l}^{~l-1/2}P_{l-1}^{1}  \notag \\
&&+\sqrt{\frac{3\left( l-1\right) }{2\left( 2l-1\right) }}\exp \left(
2i\omega _{l}\right) U_{1/2l)3/2l}^{~l-1/2}P_{l}^{1}  \notag \\
&&-\sqrt{\frac{\left( l-2\right) }{2\left( 2l-1\right) }}\exp \left( i\left(
\omega _{l}+\omega _{l-2}\right) \right) U_{1/2(l-2)3/2l}^{~l-3/2}P_{l-2}^{1}
\notag \\
&&+\left. \sqrt{\frac{\left( l-2\right) }{2\left( 2l-1\right) }}\exp \left(
i\left( \omega _{l}+\omega _{l-1}\right) \right)
U_{1/2(l-1)3/2l}^{~l-3/2}P_{l-1}^{1}\right]
\end{eqnarray}

\begin{eqnarray}
\mathfrak{M}_{1/2\text{\ \ }1/2}^{3/2-3/2} &=&\frac{i}{2k}\sum\limits_{l}%
\left[ \frac{1}{\sqrt{2l\left( 2l+3\right) \left( l+1\right) }}\exp \left(
i\left( \omega _{l}+\omega _{l+1}\right) \right)
U_{1/2(l+1)3/2l}^{~l+3/2}P_{l+1}^{2}\right.  \notag \\
&&+\frac{1}{\sqrt{2l\left( 2l+3\right) \left( l+1\right) }}\exp \left(
i\left( \omega _{l}+\omega _{l+2}\right) \right)
U_{1/2(l+2)3/2l}^{~l+3/2}P_{l+2}^{2}  \notag \\
&&-\sqrt{\frac{3}{2l\left( 2l+3\right) }}\exp \left( 2i\omega _{l}\right)
U_{1/2(l)3/2l}^{~l+1/2}P_{l}^{2}  \notag \\
&&-\sqrt{\frac{3}{2l\left( 2l+3\right) }}\exp \left( i\left( \omega
_{l}+\omega _{l+1}\right) \right) U_{1/2(l+1)3/2l}^{~l+1/2}P_{l+1}^{2}
\notag \\
&&+\sqrt{\frac{3}{2\left( 2l-1\right) \left( l+1\right) }}\exp \left(
i\left( \omega _{l}+\omega _{l-1}\right) \right)
U_{1/2(l-1)3/2l}^{~l-1/2}P_{l-1}^{2}  \notag \\
&&+\sqrt{\frac{3}{2\left( 2l-1\right) \left( l+1\right) }}\exp \left(
2i\omega _{l}\right) U_{1/2l)3/2l}^{~l-1/2}P_{l}^{2}  \notag \\
&&-\frac{1}{\sqrt{2l\left( 2l-1\right) }}\exp \left( i\left( \omega
_{l}+\omega _{l-2}\right) \right) U_{1/2(l-2)3/2l}^{~l-3/2}P_{l-2}^{2}
\notag \\
&&-\left. \frac{1}{\sqrt{2l\left( 2l-1\right) }}\exp \left( i\left( \omega
_{l}+\omega _{l-1}\right) \right) U_{1/2(l-1)3/2l}^{~l-3/2}P_{l-1}^{2}\right]
\label{B22}
\end{eqnarray}

\section{\protect\bigskip Partial amplitudes for the channel spin S=5/2}

\begin{eqnarray}
\mathcal{S}_{\text{ \ \ \ \ \ }1/2}^{5/2\text{ }1/2} &=&f_{c}(\theta )+\frac{%
1}{4ik}\sum\limits_{l=0}\left[ \frac{5\left( l+1\right) \left( l+2\right)
\left( l+3\right) }{\left( 2l+3\right) \left( 2l+5\right) }\alpha ^{l+5/2}+%
\frac{l\left( l+1\right) \left( l+2\right) }{\left( 2l+3\right) \left(
2l+5\right) }\alpha ^{l+3/2}+\frac{2l\left( l+1\right) \left( l+2\right) }{%
\left( 2l+3\right) \left( 2l-1\right) }\alpha ^{l+1/2}\right.  \notag \\
&&\left. +\frac{2l\left( l+1\right) \left( l-1\right) }{\left( 2l+3\right)
\left( 2l-1\right) }\alpha ^{l-1/2}+\frac{l\left( l+1\right) \left(
l-1\right) }{\left( 2l-1\right) \left( 2l-3\right) }\alpha ^{l-3/2}+\frac{%
5l\left( l-1\right) \left( l-2\right) }{\left( 2l-3\right) \left(
2l-1\right) }\alpha ^{l-5/2}\right]  \notag \\
&&\times \exp \left( 2i\omega _{l}\right) P_{l}\left( \cos \theta \right) ,
\label{C1A6}
\end{eqnarray}

\begin{eqnarray}
\mathcal{S}_{\text{ \ \ \ \ \ }3/2}^{5/2\text{ }3/2} &=&f_{c}(\theta )+\frac{%
1}{8ik}\sum\limits_{l=0}\left[ \frac{5\left( l+1\right) \left( l+3\right)
\left( l+4\right) }{\left( 2l+3\right) \left( 2l+5\right) }\alpha ^{l+5/2}+%
\frac{9l\left( l+2\right) \left( l+3\right) }{\left( 2l+3\right) \left(
2l+5\right) }\alpha ^{l+3/2}\right.  \notag \\
&&+\frac{2l\left( l+1\right) \left( l-3\right) ^{2}}{\left( 2l+3\right)
\left( 2l-1\right) }\alpha ^{l+1/2}+\frac{2l\left( l+4\right) ^{2}}{\left(
2l+3\right) \left( 2l-1\right) }\alpha ^{l-1/2}+\frac{9\left( l+1\right)
\left( l-1\right) \left( l-2\right) }{\left( 2l-1\right) \left( 2l-3\right) }%
\alpha ^{l-3/2}  \notag \\
&&\left. +\frac{5l\left( l-2\right) \left( l-3\right) }{\left( 2l-3\right)
\left( 2l-1\right) }\alpha ^{l-5/2}\right] \exp \left( 2i\omega _{l}\right)
P_{l}\left( \cos \theta \right) ,  \label{B6}
\end{eqnarray}

\begin{eqnarray}
\mathcal{S}_{\text{ \ \ \ \ \ }5/2}^{5/2\text{ }5/2} &=&f_{c}(\theta )+\frac{%
1}{8ik}\sum\limits_{l=0}\left[ \frac{\left( l+3\right) \left( l+4\right)
\left( l+5\right) }{\left( 2l+3\right) \left( 2l+5\right) }\alpha ^{l+5/2}+%
\frac{5\left( l+2\right) \left( l+3\right) \left( l+4\right) }{\left(
2l+3\right) \left( 2l+5\right) }\alpha ^{l+3/2}\right.  \notag \\
&&+\frac{10\left( l+1\right) \left( l-1\right) \left( l+3\right) }{\left(
2l+3\right) \left( 2l-1\right) }\alpha ^{l+1/2}+\frac{10l\left( l+2\right)
\left( l-2\right) }{\left( 2l+3\right) \left( 2l-1\right) }\alpha ^{l-1/2}+%
\frac{5\left( l-1\right) \left( l-2\right) \left( l-3\right) }{\left(
2l-1\right) \left( 2l-3\right) }\alpha ^{l-3/2}  \notag \\
&&\left. +\frac{\left( l-2\right) \left( l-3\right) \left( l-4\right) }{%
\left( 2l-3\right) \left( 2l-1\right) }\alpha ^{l-5/2}\right] \exp \left(
2i\omega _{l}\right) P_{l}\left( \cos \theta \right) ,  \label{C6}
\end{eqnarray}

\begin{eqnarray}
\mathcal{S}_{\text{ \ \ \ }-1/2}^{5/2\text{ }1/2} &=&-\frac{1}{4ik}%
\sum\limits_{l=1}\left[ \frac{5\left( l+2\right) \left( l+3\right) }{\left(
2l+3\right) \left( 2l+5\right) }\alpha ^{l+5/2}-\frac{\left( l+2\right)
\left( l+5\right) }{\left( l+3\right) \left( 2l+5\right) }\alpha ^{l+3/2}+%
\frac{2\left( l+2\right) \left( l-2\right) }{\left( 2l+3\right) \left(
2l-1\right) }\alpha ^{l+1/2}\right.  \notag \\
&&\left. -\frac{2\left( l+3\right) \left( l-1\right) }{\left( 2l+3\right)
\left( 2l-1\right) }\alpha ^{l-1/2}+\frac{\left( l-1\right) \left(
l-4\right) }{\left( 2l-1\right) \left( 2l-3\right) }\alpha ^{l-3/2}-\frac{%
5\left( l-1\right) \left( l-2\right) }{\left( 2l-3\right) \left( 2l-1\right)
}\alpha ^{l-5/2}\right]  \notag \\
&&\times \exp \left( 2i\omega _{l}\right) P_{l}^{1}\left( \cos \theta
\right) ,  \label{D6}
\end{eqnarray}

\begin{eqnarray}
\mathcal{S}_{\text{ \ \ \ \ \ }3/2}^{5/2\text{ }1/2} &=&-\frac{\sqrt{2}}{8ik}%
\sum\limits_{l=1}\left[ \frac{5\left( l+2\right) \left( l+3\right) }{\left(
2l+3\right) \left( 2l+5\right) }\alpha ^{l+5/2}+\frac{\left( l+2\right)
\left( 3l+5\right) }{\left( 2l+3\right) \left( 2l+5\right) }\alpha ^{l+3/2}-%
\frac{2\left( l+2\right) \left( l+1\right) }{\left( 2l+3\right) \left(
2l-1\right) }\alpha ^{l+1/2}\right.  \notag \\
&&\left. +\frac{2l\left( l-1\right) }{\left( 2l+3\right) \left( 2l-1\right) }%
\alpha ^{l-1/2}-\frac{\left( l-1\right) \left( 3l-2\right) }{\left(
2l-1\right) \left( 2l-3\right) }\alpha ^{l-3/2}-\frac{5\left( l-1\right)
\left( l-2\right) }{\left( 2l-3\right) \left( 2l-1\right) }\alpha ^{l-5/2}%
\right]  \notag \\
&&\times \exp \left( 2i\omega _{l}\right) P_{l}^{1}\left( \cos \theta
\right) ,  \label{E6}
\end{eqnarray}

\begin{eqnarray}
\mathcal{S}_{\text{ \ \ \ \ \ }1/2}^{5/2\text{ }3/2} &=&-\frac{\sqrt{2}}{8ik}%
\sum\limits_{l=1}\left[ \frac{5\left( l+3\right) \left( l+4\right) }{\left(
2l+3\right) \left( 2l+5\right) }\alpha ^{l+5/2}+\frac{3\left( l+2\right)
\left( l+3\right) \left( l-5\right) }{\left( l+1\right) \left( 2l+3\right)
\left( 2l+5\right) }\alpha ^{l+3/2}-\frac{2\left( l^{2}+4l-6\right) \left(
l-3\right) }{l\left( 2l+3\right) \left( 2l-1\right) }\alpha ^{l+1/2}\right.
\notag \\
&&\left. +\frac{2l\left( l+4\right) \left( l^{2}-2l-9\right) }{\left(
l+1\right) \left( 2l+3\right) \left( 2l-1\right) }\alpha ^{l-1/2}-\frac{%
3\left( l+6\right) \left( l-1\right) \left( l-2\right) }{l\left( 2l-1\right)
\left( 2l-3\right) }\alpha ^{l-3/2}-\frac{5\left( l-2\right) \left(
l-3\right) }{\left( 2l-3\right) \left( 2l-1\right) }\alpha ^{l-5/2}\right]
\notag \\
&&\times \exp \left( 2i\omega _{l}\right) P_{l}^{1}\left( \cos \theta
\right) ,  \label{F6}
\end{eqnarray}

\begin{eqnarray}
\mathcal{S}_{\text{ \ \ \ \ \ }5/2}^{5/2\text{ }3/2} &=&-\frac{\sqrt{5}}{8ik}%
\sum\limits_{l=1}\left[ \frac{\left( l+3\right) \left( l+4\right) }{\left(
2l+3\right) \left( 2l+5\right) }\alpha ^{l+5/2}+\frac{3\left( l+2\right)
\left( l+3\right) }{\left( 2l+3\right) \left( 2l+5\right) }\alpha ^{l+3/2}+%
\frac{2\left( l+1\right) \left( l-3\right) }{\left( 2l+3\right) \left(
2l-1\right) }\alpha ^{l+1/2}\right.  \notag \\
&&\left. -\frac{2l\left( l+4\right) }{\left( 2l+3\right) \left( 2l-1\right) }%
\alpha ^{l-1/2}-\frac{3\left( l-1\right) \left( l-2\right) }{\left(
2l-1\right) \left( 2l-3\right) }\alpha ^{l-3/2}-\frac{\left( l-2\right)
\left( l-3\right) }{\left( 2l-3\right) \left( 2l-1\right) }\alpha ^{l-5/2}%
\right]  \notag \\
&&\times \exp \left( 2i\omega _{l}\right) P_{l}^{1}\left( \cos \theta \right)
\end{eqnarray}

\begin{eqnarray}
\mathcal{S}_{\text{ \ \ \ \ \ }3/2}^{5/2\text{ }5/2} &=&-\frac{\sqrt{5}}{8ik}%
\sum\limits_{l=1}\left[ \frac{\left( l+3\right) \left( l+4\right) \left(
l+5\right) }{\left( l+1\right) \left( 2l+3\right) \left( 2l+5\right) }\alpha
^{l+5/2}+\frac{\left( l+2\right) \left( l+3\right) \left( l+4\right) \left(
3l-5\right) }{l\left( l+1\right) \left( 2l+3\right) \left( 2l+5\right) }%
\alpha ^{l+3/2}\right.  \notag \\
&&+\frac{2\left( l+3\right) \left( l-1\right) \left( l-8\right) }{l\left(
2l+3\right) \left( 2l-1\right) }\alpha ^{l+1/2}-\frac{2\left( l+9\right)
\left( l+2\right) \left( l-2\right) }{\left( l+1\right) \left( 2l+3\right)
\left( 2l-1\right) }\alpha ^{l-1/2}-\frac{\left( 3l+8\right) \left(
l-1\right) \left( l-2\right) \left( l-3\right) }{l\left( l+1\right) \left(
2l-1\right) \left( 2l-3\right) }\alpha ^{l-3/2}  \notag \\
&&\left. -\frac{\left( l-2\right) \left( l-3\right) \left( l-4\right) }{%
l\left( 2l-3\right) \left( 2l-1\right) }\alpha ^{l-5/2}\right] \exp \left(
2i\omega _{l}\right) P_{l}^{1}\left( \cos \theta \right) ,  \label{H6}
\end{eqnarray}

\begin{eqnarray}
\mathcal{S}_{\text{ \ \ \ \ \ \ }3/2}^{5/2-1/2} &=&\frac{\sqrt{2}}{8ik}%
\sum\limits_{l=2}\left[ \frac{5\left( l+3\right) }{\left( 2l+3\right) \left(
2l+5\right) }\alpha ^{l+5/2}-\frac{3l+10}{\left( 2l+3\right) \left(
2l+5\right) }\alpha ^{l+3/2}-\frac{2\left( l+7\right) }{\left( 2l+3\right)
\left( 2l-1\right) }\alpha ^{l+1/2}\right.  \notag \\
&&\left. -\frac{2\left( l-6\right) }{\left( 2l+3\right) \left( 2l-1\right) }%
\alpha ^{l-1/2}-\frac{3l-7}{\left( 2l-1\right) \left( 2l-3\right) }\alpha
^{l-3/2}+\frac{5\left( l-2\right) }{\left( 2l-3\right) \left( 2l-1\right) }%
\alpha ^{l-5/2}\right]  \notag \\
&&\times \exp \left( 2i\omega _{l}\right) P_{l}^{2}\left( \cos \theta
\right) ,  \label{I6}
\end{eqnarray}

\begin{eqnarray}
\mathcal{S}_{\text{ \ \ \ \ \ }5/2}^{5/2\text{ }1/2} &=&\frac{\sqrt{10}}{8ik}%
\sum\limits_{l=2}\left[ \frac{l+3}{\left( 2l+3\right) \left( 2l+5\right) }%
\alpha ^{l+5/2}+\frac{l+2}{\left( 2l+3\right) \left( 2l+5\right) }\alpha
^{l+3/2}-\frac{2\left( l+1\right) }{\left( 2l+3\right) \left( 2l-1\right) }%
\alpha ^{l+1/2}\right.  \notag \\
&&\left. -\frac{2l}{\left( 2l+3\right) \left( 2l-1\right) }\alpha ^{l-1/2}+%
\frac{l-1}{\left( 2l-1\right) \left( 2l-3\right) }\alpha ^{l-3/2}+\frac{l-2}{%
\left( 2l-3\right) \left( 2l-1\right) }\alpha ^{l-5/2}\right]  \notag \\
&&\times \exp \left( 2i\omega _{l}\right) P_{l}^{2}\left( \cos \theta
\right) ,  \label{J6}
\end{eqnarray}

\begin{eqnarray}
\mathcal{S}_{\text{ \ \ \ \ \ \ }1/2}^{5/2-3/2} &=&\frac{1}{4\sqrt{2}ik}%
\sum\limits_{l=2}\left[ \frac{5\left( l+3\right) \left( l+4\right) }{\left(
l+2\right) \left( 2l+3\right) \left( 2l+5\right) }\alpha ^{l+5/2}-\frac{%
3\left( l+3\right) \left( l+10\right) }{\left( l+1\right) \left( 2l+3\right)
\left( 2l+5\right) }\alpha ^{l+3/2}\right.  \notag \\
&&-\frac{2\left( l-3\right) \left( l^{2}-2l-18\right) }{l\left( l+2\right)
\left( 2l+3\right) \left( 2l-1\right) }\alpha ^{l+1/2}-\frac{2\left(
l+4\right) \left( l^{2}+4l-15\right) }{\left( l+1\right) \left( l-1\right)
\left( 2l+3\right) \left( 2l-1\right) }\alpha ^{l-1/2}-\frac{3\left(
l-2\right) \left( l-9\right) }{l\left( 2l-1\right) \left( 2l-3\right) }%
\alpha ^{l-3/2}  \notag \\
&&\left. +\frac{5\left( l-2\right) \left( l-3\right) }{\left( l-1\right)
\left( 2l-3\right) \left( 2l-1\right) }\alpha ^{l-5/2}\right] \exp \left(
2i\omega _{l}\right) P_{l}^{2}\left( \cos \theta \right) ,  \label{K6}
\end{eqnarray}

\begin{eqnarray}
\mathcal{S}_{\text{ \ \ \ \ \ }1/2}^{5/2\text{ }5/2} &=&\frac{\sqrt{10}}{8ik}%
\sum\limits_{l=2}\left[ \frac{\left( l+3\right) \left( l+4\right) \left(
l+5\right) }{\left( l+1\right) \left( l+2\right) \left( 2l+3\right) \left(
2l+5\right) }\alpha ^{l+5/2}+\frac{\left( l+3\right) \left( l+4\right)
\left( l-10\right) }{l\left( l+1\right) \left( 2l+3\right) \left(
2l+5\right) }\alpha ^{l+3/2}\right.  \notag \\
&&-\frac{2\left( l+3\right) \left( l^{2}+6l-22\right) }{l\left( l+2\right)
\left( 2l+3\right) \left( 2l-1\right) }\alpha ^{l+1/2}-\frac{2\left(
l-2\right) \left( l^{2}-4l-27\right) }{\left( l+1\right) \left( l-1\right)
\left( 2l+3\right) \left( 2l-1\right) }\alpha ^{l-1/2}+\frac{\left(
l-2\right) \left( l-3\right) \left( l+11\right) }{l\left( l+1\right) \left(
2l-1\right) \left( 2l-3\right) }\alpha ^{l-3/2}  \notag \\
&&\left. +\frac{\left( l-2\right) \left( l-3\right) \left( l-4\right) }{%
l\left( l-1\right) \left( 2l-3\right) \left( 2l-1\right) }\alpha ^{l-5/2}%
\right] \exp \left( 2i\omega _{l}\right) P_{l}^{2}\left( \cos \theta \right)
,  \label{L6}
\end{eqnarray}

\begin{eqnarray}
\mathcal{S}_{\text{ \ \ \ }-5/2}^{5/2\text{ }1/2} &=&-\frac{\sqrt{10}}{8ik}%
\sum\limits_{l=3}\left[ \frac{1}{\left( 2l+3\right) \left( 2l+5\right) }%
\alpha ^{l+5/2}-\frac{1}{\left( 2l+3\right) \left( 2l+5\right) }\alpha
^{l+3/2}-\frac{2}{\left( 2l+3\right) \left( 2l-1\right) }\alpha
^{l+1/2}\right.  \notag \\
&&\left. +\frac{2}{\left( 2l+3\right) \left( 2l-1\right) }\alpha ^{l-1/2}+%
\frac{1}{\left( 2l-1\right) \left( 2l-3\right) }\alpha ^{l-3/2}-\frac{1}{%
\left( 2l-3\right) \left( 2l-1\right) }\alpha ^{l-5/2}\right]  \notag \\
&&\times \exp \left( 2i\omega _{l}\right) P_{l}^{3}\left( \cos \theta
\right) ,  \label{M6}
\end{eqnarray}

\begin{eqnarray}
\mathcal{S}_{\text{ \ \ \ }-3/2}^{5/2\text{ }3/2} &=&-\frac{1}{8ik}%
\sum\limits_{l=3}\left[ \frac{5\left( l+4\right) }{\left( l+2\right) \left(
2l+3\right) \left( 2l+5\right) }\alpha ^{l+5/2}-\frac{9\left( l+5\right) }{%
\left( l+1\right) \left( 2l+3\right) \left( 2l+5\right) }\alpha
^{l+3/2}\right.  \notag \\
&&+\frac{2\left( l+12\right) \left( l-3\right) }{l\left( l+2\right) \left(
2l+3\right) \left( 2l-1\right) }\alpha ^{l+1/2}-\frac{2\left( l+4\right)
\left( l-11\right) }{\left( l+1\right) \left( l-1\right) \left( 2l+3\right)
\left( 2l-1\right) }\alpha ^{l-1/2}+\frac{9\left( l-4\right) }{l\left(
2l-1\right) \left( 2l-3\right) }\alpha ^{l-3/2}  \notag \\
&&\left. -\frac{5\left( l-3\right) }{\left( l-1\right) \left( 2l-3\right)
\left( 2l-1\right) }\alpha ^{l-5/2}\right] \exp \left( 2i\omega _{l}\right)
P_{l}^{3}\left( \cos \theta \right) ,  \label{N6}
\end{eqnarray}

\begin{eqnarray}
\mathcal{S}_{\text{ \ \ \ }-1/2}^{5/2\text{ }5/2} &=&-\frac{\sqrt{10}}{8ik}%
\sum\limits_{l=3}\left[ \frac{\left( l+4\right) \left( l+5\right) }{\left(
l+1\right) \left( l+2\right) \left( 2l+3\right) \left( 2l+5\right) }\alpha
^{l+5/2}-\frac{\left( l+4\right) \left( l+15\right) }{l\left( l+1\right)
\left( 2l+3\right) \left( 2l+5\right) }\alpha ^{l+3/2}\right.  \notag \\
&&-\frac{2\left( l^{2}-4l-42\right) }{l\left( l+2\right) \left( 2l+3\right)
\left( 2l-1\right) }\alpha ^{l+1/2}+\frac{2\left( l^{2}+6l-37\right) }{%
\left( l+1\right) \left( l-1\right) \left( 2l+3\right) \left( 2l-1\right) }%
\alpha ^{l-1/2}+\frac{\left( l-3\right) \left( l-14\right) }{l\left(
l+1\right) \left( 2l-1\right) \left( 2l-3\right) }\alpha ^{l-3/2}  \notag \\
&&\left. -\frac{\left( l-3\right) \left( l-4\right) }{l\left( l-1\right)
\left( 2l-3\right) \left( 2l-1\right) }\alpha ^{l-5/2}\right] \exp \left(
2i\omega _{l}\right) P_{l}^{3}\left( \cos \theta \right) ,  \label{O6}
\end{eqnarray}

\begin{eqnarray}
\mathcal{S}_{\text{ \ \ \ \ \ \ }5/2}^{5/2-3/2} &=&\frac{5}{8ik}%
\sum\limits_{l=4}\left[ \frac{1}{\left( l+2\right) \left( 2l+3\right) \left(
2l+5\right) }\alpha ^{l+5/2}-\frac{3}{\left( l+1\right) \left( 2l+3\right)
\left( 2l+5\right) }\alpha ^{l+3/2}+\frac{2\left( l-3\right) }{l\left(
l+2\right) \left( 2l+3\right) \left( 2l-1\right) }\alpha ^{l+1/2}\right.
\notag \\
&&\left. +\frac{2\left( l+4\right) }{\left( l+1\right) \left( l-1\right)
\left( 2l+3\right) \left( 2l-1\right) }\alpha ^{l-1/2}-\frac{3}{l\left(
2l-1\right) \left( 2l-3\right) }\alpha ^{l-3/2}+\frac{1}{\left( l-1\right)
\left( 2l-3\right) \left( 2l-1\right) }\alpha ^{l-5/2}\right]  \notag \\
&&\times \exp \left( 2i\omega _{l}\right) P_{l}^{4}\left( \cos \theta
\right) ,  \label{P6}
\end{eqnarray}

\begin{eqnarray}
\mathcal{S}_{\text{ \ \ \ \ \ \ }3/2}^{5/2-5/2} &=&\frac{5}{8ik}%
\sum\limits_{l=4}\left[ \frac{l+5}{\left( l+1\right) \left( l+2\right)
\left( 2l+3\right) \left( 2l+5\right) }\alpha ^{l+5/2}-\frac{3l+20}{l\left(
l+1\right) \left( 2l+3\right) \left( 2l+5\right) }\alpha ^{l+3/2}\right.
\notag \\
&&+\frac{2\left( l+17\right) }{l\left( l+2\right) \left( 2l+3\right) \left(
2l-1\right) }\alpha ^{l+1/2}+\frac{2\left( l-16\right) }{\left( l+1\right)
\left( l-1\right) \left( 2l+3\right) \left( 2l-1\right) }\alpha ^{l-1/2}-%
\frac{3l-17}{l\left( l+1\right) \left( 2l-1\right) \left( 2l-3\right) }%
\alpha ^{l-3/2}  \notag \\
&&\left. +\frac{l-4}{l\left( l-1\right) \left( 2l-3\right) \left(
2l-1\right) }\alpha ^{l-5/2}\right] \exp \left( 2i\omega _{l}\right)
P_{l}^{4}\left( \cos \theta \right) ,  \label{Q6}
\end{eqnarray}

\begin{equation}
\mathcal{S}_{\text{ \ \ \ }-5/2}^{5/2\text{ }5/2}=-\frac{1}{2ik}%
\sum\limits_{l=5}\sqrt{\frac{\left( l-5\right) !}{\left( l+5\right) !}}\left[
\alpha ^{l+5/2}-5\alpha ^{l+3/2}+10\alpha ^{l+1/2}-10\alpha ^{l-1/2}+5\alpha
^{l-3/2}-\alpha ^{l-5/2}\right] \exp \left( 2i\omega _{l}\right)
P_{l}^{5}\left( \cos \theta \right) .  \label{C18R6}
\end{equation}

\end{document}